\documentclass[journal=nalefd,manuscript=letter]{achemso}

\usepackage{graphicx}
\usepackage{dcolumn}
\usepackage{bm}
\usepackage{graphicx}
\usepackage{subfig}
\usepackage{comment}
\usepackage{amsmath}
\usepackage{amssymb}
\usepackage{soul}  
\usepackage[below]{placeins}
\usepackage[usenames, dvipsnames]{color}
\usepackage{siunitx}
\usepackage{float}
\usepackage{tabularx}
\usepackage{hyperref}
\usepackage{cleveref}
\usepackage{pdfpages}
\usepackage{lipsum}
\usepackage{booktabs} 
\usepackage[font={small}]{caption}
\usepackage[mathscr]{eucal}
\DeclareMathAlphabet{\mathpzc}{OT1}{pzc}{m}{it}

\usepackage{lipsum} 
\usepackage{comment}

\usepackage[labelfont=bf, labelsep=period]{caption}
\hypersetup{
colorlinks   = true, 
  urlcolor     = blue, 
  linkcolor    = black, 
  citecolor   = black 
}

\title{Spin and Orbital Angular Momentum Polarization in Thouless Topological Charge Pumping}

\author{Esmaeil Taghizadeh Sisakht}
\affiliation{Department of Physics, Ulsan National Institute of Science and Technology, Ulsan, 689-798, Korea}

\author{Uiseok Jeong}
\affiliation{Department of Physics, Ulsan National Institute of Science and Technology, Ulsan, 689-798, Korea}
\author{Xiao jiang}
\affiliation{Department of Physics, Ulsan National Institute of Science and Technology, Ulsan, 689-798, Korea}
\author{Jinseok Oh}
   \affiliation{Department of Physics, Ulsan National Institute of Science and Technology, Ulsan, 689-798, Korea}
\author{Yizhou Liu}
\affiliation{Center for Phononics and Thermal Energy Science, China-EU Joint Lab on Nanophononics,
School of Physics Science and Engineering, Tongji University, 200092 Shanghai, China}

 \author{Binghai Yan}
 \email{binghai.yan@psu.edu }
   \affiliation{Department of Physics, Pennsylvania State University,
201 Old Main, University Park, Pennsylvania, 16802, USA}

 \author{Noejung Park}
   \affiliation{Department of Physics, Ulsan National Institute of Science and Technology, Ulsan, 689-798, Korea}
    \email{noejung@unist.ac.kr}

\begin{document}



\begin{abstract}
Quantized charge pumping in one-dimensional chiral wires has been widely studied in
the context of topological physics in a (1+1)-dimensional synthetic space, yet the role of
orbital and spin degrees of freedom in such topological pumps remains largely
unexplored. Here, we examine how the topologically quantized charge pump in
insulators generates spin polarizations, and assess whether this mechanism may offer
distinct insight into the widely known spin-selective transport in chiral wires—commonly
referred to as chirality-induced spin selectivity. We performed time-dependent
Schrödinger equations of multi-orbital tight-binding Hamiltonians driven by a circularly
polarized electric field. Our main findings are twofold.
First, the intrinsic screw-like geometry of the system generates a distinctive winding structure governed by a single control parameter, in contrast to conventional adiabatic pumping mechanisms that require at least two independently modulated parameters, thereby providing a clear interpretation of one-dimensional pumping in terms of the topological structure in a (1+1)-dimensional Brillouin zone.
Second, while the energy gap remains open
throughout the pumping cycle, the Berry-phase driven real-time dynamics of the charge
flow induces a nonequilibrium orbital polarization. Through spin-orbit coupling, this
orbital response is partially converted into spin polarization whose direction is determined by the
current and chirality. On the analogy between the synthetic (1+1)- and 2-dimensional
topological insulators, we suggest that non-trivial spin-orbital dynamics may accompany
the anomalous quantum charge Hall states of even-dimensional real materials.
\end{abstract}
%
%
\section{Introduction}
The quantum pump, as a dynamical
manifestation of topological phases, has attracted significant interest for its fundamental
physics and perspective applications of topological charge transport ~\cite{laughlin1981quantized,thouless1983quantization,wang2013topological,nakajima2016topological,lohse2016thouless,nakajima2021competition,sun2022non,grinberg2020robust,chen2021topological,song2024fast,wang2022two,haug2019topological}. The underlying theoretical foundations
can be traced back to Laughlin, whose explanation of the quantum Hall effect in terms of a
cylindrical geometry with a varying magnetic flux directly inspired the concept of topological
pumping~\cite{laughlin1981quantized}. 
Similarly, as demonstrated by Thouless, 
the quantized charge transport in a cyclic one-dimensional (1D) Hilbert space can be mathematically mapped to the two-dimensional (2D) 
 quantized Hall conductance~\cite{thouless1983quantization}.
Beyond this mathematical analogy between 1D
and 2D, efforts have been made to realize the Thouless pump by driving a one-dimensional
Hamiltonian with a time-periodic potential, thereby forming an effective (1+1)-dimensional
system that uses time as a synthetic second dimension~\cite{wang2013topological,nakajima2016topological}. In this context, various
experimental platforms have been explored, including photonic, magneto-mechanical,
electro-mechanical, acoustic, and ultracold atom systems~\cite{nakajima2016topological,lohse2016thouless,nakajima2021competition,sun2022non,grinberg2020robust,chen2021topological,song2024fast}.  In pure condensed
matter settings, however, the one-dimensional charge pump has not yet been well realized
experimentally and has largely remained at the level of model Hamiltonians, such as the
Rice-Mele model~\cite{rice1982elementary}. On the road toward the real-material realization of topological
Thouless charge pumping, valuable clues can be found in previous studies of one-
dimensional chiral materials, such as DNA and $\alpha$-helical proteins, which have been shown to
host topological states when subjected to a low-frequency rotating   electric
field~\cite{guo2017topological,guo2020topological}.\\
Among the chirality-driven phenomena that
have recently attracted significant attention, a particularly
intriguing example  is chirality-induced spin selectivity
(CISS)~\cite{naaman2019chiral, evers2022theory,adhikari2023interplay,abendroth2019spin,guo2014spin}. This effect is observed in various molecular and solid-state systems, in which electrons get spin-polarized
as they travel through the chiral structure. While spin-selective
currents are typically associated with magnetic materials or systems with strong spin-orbit
coupling (SOC), the spin polarization mediated by the CISS mechanism emerges in chiral
materials composed primarily of light atoms with weak local SOC. Although the sound
foundation for understanding the phenomenon is still debated, it is widely accepted that the
physical origin of the CISS effect originates from the coupling between the electron
momentum and spin due to the symmetry breaking imposed by the chiral structure~\cite{bloom2024chiral}. This underscores the unexpectedly large momentum-orbital-spin coupling in CISS-active materials,
as it plays a crucial role in facilitating this spin-momentum coupling.
In this context, the   orbital angular momentum (OAM) polarization which can develop even in materials composed of light atoms, has increasingly been recognized as a fundamental ingredient in understanding these phenomena.
\\
In this work, using dynamical time-dependent tight-binding (TB) simulations, we demonstrate that quantized Thouless charge pumping can be realized in real chiral materials, focusing on two representative systems: a spinless chiral hydrocarbon C$_{12}$H$_{12}$ and a trigonal chiral wire with intrinsic spin-orbit interaction.
We show that the intrinsic chiral feature of the wire impose a distinct
winding structure of the Bloch states from the conventional two-parameter Rice-Mele
pump, thereby reducing the realization of topological Thouless pumping to a single control parameter. In the quantum regime, the sliding potential minimum generated by a circularly polarized drive not only transports a localized particle but also imprints a geometric phase on its wavefunction. Over a full modulation cycle, this accumulated Berry phase leads to a quantized change in the electronic polarization along the wire, which is topologically equivalent to a Chern number in the synthetic (1+1)-dimensional synthetic space.
We next address the role of orbital and spin degrees of freedom in such topological pumps. We show that the resulting topological charge transport is accompanied by a finite nonequilibrium orbital polarization, arising from the momentum-dependent orbital texture of the chiral band structure. Through the local SOC, this orbital response is partially converted into spin angular momentum (SAM) polarization, providing a purely topological and all-electric route to CISS in gapped chiral systems.
Our findings establish chiral wires as realistic platforms for orbital- and spin-polarized Thouless charge pumping, and motivate a broader search for Berry-phase-driven spin-orbital textures in chiral quantum materials and equivalent anomalous Hall systems.
\section{Results}
\begin{figure}[t!]
    \centering
    \includegraphics[width=.75\textwidth]{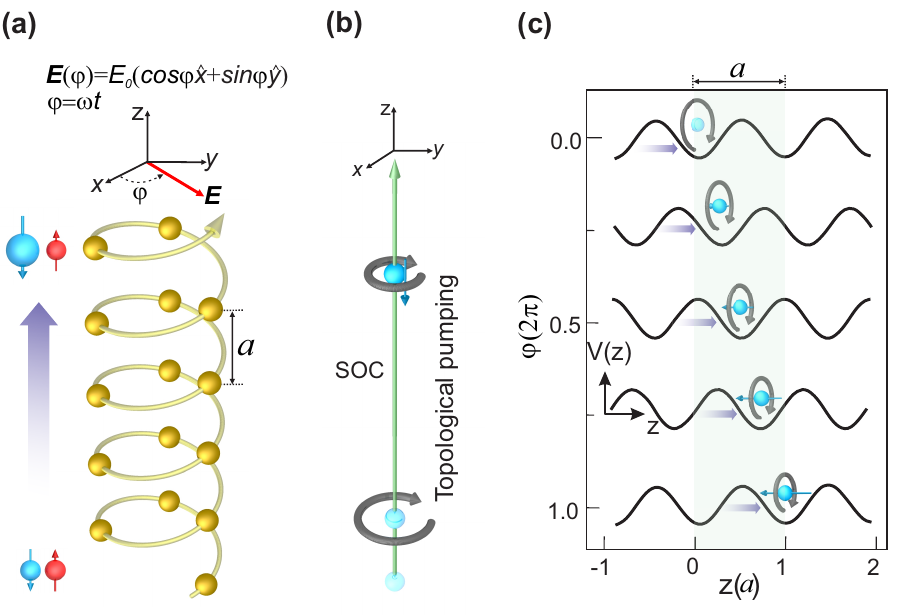} 
\caption{{\bf $\bm{|}$  Orbital-spin polarized topological charge pumping in a chiral single wire.} \textbf{a},  Schematic of chiral-induced  spin polarization via topological charge pumping. A low-frequency, circularly polarized electric field $\bm{E}$ is applied   to the chiral wire to act as the adiabatic drive.
\textbf{b,} As the electron is pumped along the chiral geometry, a nonequilibrium orbital polarization $L_z$ is generated, which is partially converted into spin through the spin-orbit interaction. 
\textbf{b,}
Illustration of the adiabatic modulation of a spatially and temporally varying potential along a 1D chiral wire, shown at several instantaneous moments.
Arrowed spheres in \textbf{a},  \textbf{b}, and \textbf{c} indicate the spin of the electron. The size of the sphere in (\textbf{a}) represents the magnitude of each spin. 
}	    
 \label{fig:fig1} 
\end{figure}

Adiabatic modulation of the energy profile along a 1D chiral geometry  provides a natural route for achieving directional transport of particles (see schematic Fig.~\ref{fig:fig1}). In the quantum dynamics, the corresponding sliding potential minimum not only transports a trapped particle but also allows its wavefunction to acquire a geometric phase. As shown by Thouless~\cite{thouless1983quantization}, the accumulated phase over a full cycle leads to quantized transport, with the pumped charge determined by topological singularities enclosed in the parameter space. Based on this idea, many proposed topological pumps utilize time-dependent superlattices to simulate the Rice-Mele model~\cite{lohse2016thouless,nakajima2016topological,lu2016geometrical,walter2023quantization}, which requires at least two sets of independently tunable parameters. Such modulation schemes are typically incompatible with solid-state systems, where the lattice structure is rigid and dynamic control demands complex experimental architectures. Moreover, the roles of orbital and spin degrees of freedom in this context have remained largely unexplored.\\
\noindent Here, we investigate chiral single-wire systems driven by a circularly polarized electric field oscillating at adiabatic frequencies (Fig.~\ref{fig:fig1}a). We demonstrate that topological Thouless pumping can be realized in such a real material using a single periodic parameter, enabled by a fundamentally distinct topological winding structure compared to the Rice-Mele model. This modulation defines a synthetic (1+1)-dimensional space and preserves a finite energy gap throughout the full cycle due to the characteristic screw-like geometry of the system and symmetry considerations. The corresponding first Chern number in this synthetic space captures the effective winding number enclosed by the trajectory in the mentioned parameter space. Moreover, we reveal that the robust topological charge pumping is accompanied by dynamical orbital polarization, which arises from the coupling between the chiral geometry and the external polarized field (Fig.~\ref{fig:fig1}b,c).  We then show that the intrinsic SOC of the chiral system further acts as a transducer, converting this orbital flow into spin-polarized transport, thereby realizing a non-magnetic mode of spin transport in non-trivial insulator chiral wires(Fig.~\ref{fig:fig1}a-c).\\
\noindent\textbf{Electronic properties of 1D chiral  wires.}
Chiral structures are known to host nontrivial momentum-dependent  orbital and spin  textures in their electronic band structures~\cite{liu2021chirality,kim2023optoelectronic}. 
To capture the essential features of the electronic structure and the orbital-spin textures of a chiral wire, we employ a three-orbital TB model  in the basis $\{ \uparrow, \downarrow \}\otimes \{ A_1, A_2, A_3,\ldots \,A_N \} \otimes \{ p_x, p_y, p_z \} $, 
where $N$ denotes the number of atoms in the unit cell.
 We  introduce the reference screw-symmetric TB Hamiltonian $\hat H_0$,  including the  local atomic SOC as (see  Supplementary Note 1 for the details):
\begin{equation}
\hat H_0 = \sum_{ij,\mu\nu,\alpha} t_{i\mu,j\nu} c^\dagger_{i\mu\alpha} c_{j\nu\alpha} 
+ \lambda_{\mathrm{SOC}} \sum_{i,\mu\nu,\alpha\beta} 
c^\dagger_{i\mu\alpha} \big( \hat{\bm{\sigma}}^{\alpha\beta}\cdot \hat{\bm{L}}^{i,\mu\nu} \big) c_{i\nu\beta}.
\label{eq:eq1}
\end{equation}
Here,   $t_{i\mu,j\nu}$  denotes the spin-independent  hopping amplitude
between the basis orbitals,  and
\( \hat{\bm{\sigma}}^{\alpha\beta} \) are the vector of Pauli matrices acting in the spin space. Also,  $\hat{\bm{L}}^{i,\mu\nu}$  represents the OAM matrix elements (in $\hbar$ units) between orbitals \( \mu \) and \( \nu \) on the same site \( i \), and $\lambda_{\mathrm{SOC}}$ is the SOC strength.\\
We first demonstrate that our TB Hamiltonian captures the salient features of the orbital and spin angular momentum textures in such  systems, which are dictated by symmetry constraints and are independent of the detailed band dispersion. In the next subsection, we analyze the symmetry-protected degeneracies in the band structure  and focus on their topological properties.
To proceed, we express the Hamiltonian in the $k$-space (Supplementary Note 1) by using the representation of the Bloch state in terms of position-space localized orbitals: \( | \chi_{\mu\sigma;k} \rangle 
= \tfrac{1}{\sqrt{N}} \sum_{\mathbf{R}} 
e^{i k \cdot (\mathbf{R}+\boldsymbol{\mu})} 
| \mathbf{R}, \boldsymbol{\mu}, \sigma \rangle \),
where \( |\mathbf{R},\boldsymbol{\mu},\sigma\rangle \) denotes the $\mu$-th orbital of 
spin \(\sigma\) at the unit cell of \(\mathbf{R}\). Using Eq.~(S7), the TB Hamiltonian in the momentum space can be written in the compact form

\begin{align}
\hat H_0(k)
&= \hat{I}_{2\times2}^{(\mathrm{spin})}\!\otimes\!\hat{T(k)}+\lambda_{\mathrm{SOC}}\,\hat{\bm{\sigma}}\cdot\!\hat{\mathbf{L}}
\label{eq:eq1k} ,
\end{align}
where \(\hat{T}(k)\) represents the spin-independent hopping  matrix.
For a 1D  wire with an $N$-fold screw symmetry,  the corresponding screw operator is defined as
\begin{equation}
\hat{\mathcal{Q}}_N = \hat{T}_z(a/N)\hat{R}_z(2\pi/N),
\label{eq:eq2}
\end{equation}
which combines a translation by $a/N$ along the chiral axis with a rotation of $\theta=2\pi/N$ about the $z$ axis.
The fractional translation \(\hat{T}_z(a/N)\) shifts orbitals along the $z$ axis and gives  
$\hat{T}_z\!\left(\tfrac{a}{N}\right)|\chi_{\mu\sigma;k}\rangle 
= e^{i k a / N} |\chi_{\mu\sigma;k}\rangle$. The Bloch eigenstates denoted by \( |\psi_{n,k}\rangle \), then inherit the same transformation property (see Supplementary Note 2).  
Since the Hamiltonian Eq.~(\ref{eq:eq1k}) commutes with the screw operator $\hat{\mathcal{Q}}_N$, the Bloch eigenstates are simultaneous eigenstates of $\hat{\mathcal{Q}}_N$. Within the atomic center approximation (ACA) (Eq.~(S6)), the expectation values of the orbital ladder operators $\hat{L}_{\pm}=\hat{L}_x \pm i\hat{L}_y$ therefore reduce to
\begin{equation}
\langle \psi_{n,k} | \hat{L}_{\pm} | \psi_{n,k} \rangle
= \langle \psi_{n,k} | \hat{R}_z^{-1}(2\pi/N)\,\hat{L}_{\pm}\,\hat{R}_z(2\pi/N) | \psi_{n,k} \rangle ,
\label{eq:eq3}
\end{equation}
where a rotation about the $z$ axis by $\theta=2\pi/N$ is generated by $\hat R_z(\theta)=\exp[-\tfrac{i\theta}{\hbar}\hat L_z]$.
The Baker-Campbell-Hausdorff relation then yields  
\begin{equation}
\langle \psi_{n,k} | \hat{L}_{\pm} | \psi_{n,k} \rangle
= e^{\pm i 2\pi/N}\langle \psi_{n,k} | \hat{L}_{\pm} | \psi_{n,k} \rangle ,
\label{eq:eq4}
\end{equation}
implying that for $i=x,y$ we have \(L_i(n,k)=\langle \psi_{n,k} | \hat{L}_{i} | \psi_{n,k}\rangle =0\) at a generic $k$, while $L_z(n,k)$ remains unconstrained.
An analogous argument applies to spin. The translational part of $\hat{\mathcal{Q}}_N$ acts trivially on spin, while the rotational part mixes the in-plane components (see Supplementary Note 3). Thus, for $\theta\neq 0$, screw symmetry enforces  
$S_x(n,k) = S_y(n,k) = 0,$
whereas $S_z(n,k)$ remains unconstrained, consistent with the invariance of $\hat S_z$ under $\hat R_z(\theta)$.  
Although the lack of additional lattice symmetries allows finite orbital and spin components along the screw axis, time-reversal symmetry enforces the antisymmetric momentum dependence $L_z(n,k) = -L_z(n,-k)$ and $S_z(n,k) = -S_z(n,-k)$.\\
In Supplementary Note 4 (Fig.~S1 and Fig.~S2), we present
and confirm that the electronic band structure and the spin-orbital textures
of this three-orbital TB model are farily consistent with those of \emph{ab initio} DFT calculations. \\
   \begin{figure}[t!]
    \centering
    \includegraphics[width=.8\textwidth]{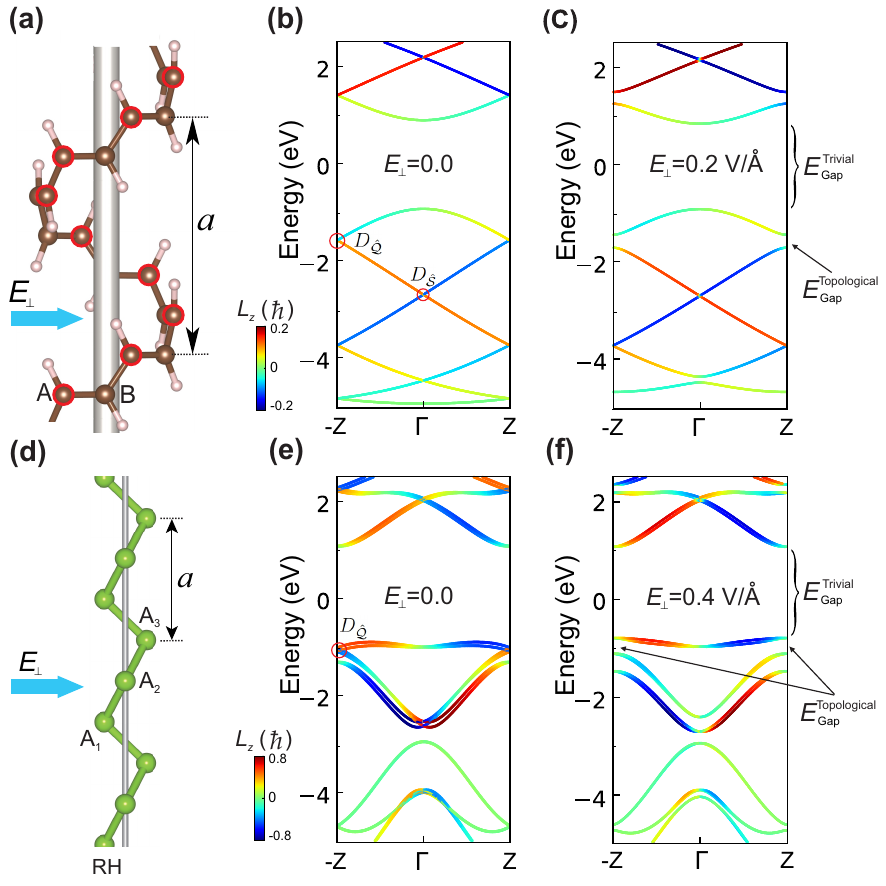} 
 \caption{{\bf $\bm{|}$ Electronic structure of chiral wires and their response to a static transverse electric field}
\textbf{a}, Atomic structure of the chiral hydrocarbon C$_{12}$H$_{12}$.  
\textbf{b},  TB band structure of C$_{12}$H$_{12}$ chiral wire at zero field, showing Dirac crossings protected by sublattice symmetry ($D_{\hat{\mathcal{S}}}$) and screw symmetry ($D_{\hat{\mathcal{Q}}}$).
\textbf{c}, Band structure of C$_{12}$H$_{12}$ chiral wire under a static transverse electric field $E_{\perp}=0.2~\mathrm{V\AA^{-1}}$, showing the field-induced gaps.  
Later, we show that the symmetry-lowering field $E_{\perp}$ drives the emergence of topological gaps $E_{\mathrm{gap}}^{\mathrm{Topological}}$, while the material itself is featured by a large trivial gap $E_{\mathrm{gap}}^{\mathrm{Trivial}}$.
\textbf{d}, Atomic structure of a trigonal  chiral wire.  
\textbf{e}, TB band structure of trigonal selenium wire including SOC at zero field.$D_{\hat{\mathcal{Q}}}$ highlights the symmetry-protected crossing at the BZ boundary.  
\textbf{f}, Band structure under a static transverse electric field $E_{\perp}=0.4~\mathrm{V\AA^{-1}}$, illustrating the gap opening in the selenium chiral wire, with $E_{\mathrm{gap}}^{\mathrm{Topological}}$ and $E_{\mathrm{gap}}^{\mathrm{Trivial}}$ indicated.
}  
 \label{fig:fig2} 
\end{figure}
\noindent\textbf{Electronic properties of 1D chiral  wires under  an electric field.}
Trigonal-stacked chiral materials such as Te and Se have been widely synthesized~\cite{starkloff1978electronic,qin2020raman,qiu2022resurrection,nakayama2024observation},
and a broad variety of chiral organic molecules have also been extensively investigated~\cite{naaman2019chiral,xie2011spin,evers2022theory,adhikari2023interplay}. It is particularly noteworthy that the chiral part of the CISS experiments usually consists of DNA-like organics without SOC\cite{naaman2019chiral,xie2011spin}. 
However, the realization of topological Thouless charge pumping in such chiral structures and in particular, the resulting orbital and spin responses during topological transport remains largely unexplored.
 In the present study, to factorize out the effect of SOC, we consider chiral hydrocarbon  C$_{12}$H$_{12}$ (Fig.~\ref{fig:fig2}a) as a model chiral system 
 which enables us to investigate purely geometric contributions during topological  charge pumping. This 1D  chiral wire inherits an extended sublattice symmetry from its carbon-nanotube precursor and additionally hosts a six-fold screw symmetry, which together enforce characteristic band degeneracies in its electronic structure. As displayed in Fig.~\ref{fig:fig2}b, the orbital resolved band structure of this spinless chiral system shows the expected symmetry-protected crossings. In the TB formulation (see Supplementary Note~5),  one may adopt the basis
$
\{A_1,\dots,A_6,\; B_1,\dots,B_6\}\otimes\{p_x,p_y,p_z\},
$
(Fig.~\ref{fig:fig2}a) which makes the two underlying symmetries discussed above particularly transparent.
In this basis, the Hamiltonian can be written in the off-diagonal form \( \hat{H}=\sigma_{+}\!\otimes \hat{ H}^{AB}+\sigma_{-}\!\otimes  \hat{H}^{BA}\) where \(\sigma_{\pm}\) are the raising and lowering Pauli matrices acting on the extended sublattice space. Using the sublattice operator \(\hat{\mathcal{S}} = \sigma_{z}\) and the identity \(\sigma_{z}\sigma_{\pm}\sigma_{z} = -\,\sigma_{\pm}\), one directly obtains the sublattice-symmetry relation \(\hat{\mathcal{S}}\, \hat{H} \,\hat{\mathcal{S}}^{-1} = -\,\hat{H}\).
As shown, the crossing at the $\Gamma$ point is protected by this sublattice symmetry (as denoted by $D_{\hat{\mathcal{S}}}$ in Fig.~\ref{fig:fig2}b). In addition, the non-symmorphic six-fold screw symmetry acting within each extended sublattice enforces Dirac-type ($D_{\hat{\mathcal{Q}}}$) crossings at the Brillouin-zone boundary $\pm Z$.\\
When a static   electric field $\mathbf{E}_{\perp}$ is applied (Fig.~\ref{fig:fig2}a), its length-gauge coupling is effectively captured by the onsite dipole term $-e\,\mathbf{E}_{\perp}\cdot\hat{\mathbf{r}}$, where $\hat{\mathbf{r}}$ denotes the intra-cell position operator. Consequently, the A/B extended sublattices experience different onsite energy shifts, explicitly breaking the sublattice symmetry. Moreover, because $\hat{\mathbf{r}}$ transforms as a polar vector in real space, any in-plane rotation of the field generates a potential that does not commute with the screw operator,
$\hat{\mathcal{Q}}_N^{-1}\,\mathbf{E}_{\perp}\cdot\hat{\mathbf{r}}\,\hat{\mathcal{Q}}_N \neq \mathbf{E}_{\perp}\cdot\hat{\mathbf{r}}$.
Thus, in the modified Hamiltonian $\hat{H}$, the rotational part of the screw symmetry about the chiral axis is violated, i.e. $\hat{\mathcal{Q}}_N^{-1}\,\hat{H}\,\hat{\mathcal{Q}}_N \neq \hat{H}$, while the translational component along  the axis remains preserved. As a result, both protecting symmetries are lifted and the Dirac points evolve into finite gaps, as shown in Fig.~\ref{fig:fig2}c.\\
Our symmetry argument remains valid in the presence of SOC.  
For the trigonal selenium chiral wire (Fig.~\ref{fig:fig2}d), the non-symmorphic three-fold screw symmetry produces Dirac-type crossings at both $\Gamma$ and the zone boundary, as seen in the TB band structure with OAM resolution (Fig.~\ref{fig:fig2}e).  
Including SOC splits each band into Kramers doublets, while time-reversal symmetry preserves the crossings between these pairs at the time-reversal-invariant momenta.  
Applying a static electric field breaks the rotational part of the screw operation, reducing the space group from \(P3_{1}21\) (or \(P3_{2}21\)) to \(P1\) and the point group from \(D_{3}\) to \(C_{1}\).  
This symmetry lowering lifts the screw-protected degeneracies and opens finite gaps at the former crossing points, as shown in Fig.~\ref{fig:fig2}f.\\
In the following, we show that the field-induced gaps acquire a nontrivial topological character.  
This topology manifests as a winding of the Bloch states across the pumping cycle, enabling a finite contribution to the Thouless charge pump that is absent in the trivial gaps of the unperturbed system.\\
\noindent\textbf{Thouless charge pumping in 1D chiral  wires.}
Applying a circularly polarized electric field \(\mathbf{E}(t) = E_0 \cos\varphi(t)\,\hat{\mathbf{x}} + E_0 \sin\varphi(t)\,\hat{\mathbf{y}}\), with \(\varphi(t) = \omega t + \varphi_0\) (\(\omega\) the driving frequency and \(\varphi_0\) the initial phase), modifies the reference Hamiltonian $\hat{H}_0$ (Eq.~\ref{eq:eq1})  and introduces a parametric dependence on \(\varphi\) in the TB Hamiltonian:  
\begin{align}
\hat{H}[\varphi] = \hat{H}_0
- e \sum_{i,\mu,\alpha} \big( \mathbf{E}(\varphi) \cdot \mathbf{r}_i \big) 
c^\dagger_{i\mu\alpha} c_{i\mu\alpha}.
\label{eq:eq6_1}
\end{align}
 \begin{figure}[t!]
    \centering
    \includegraphics[width=1\textwidth]{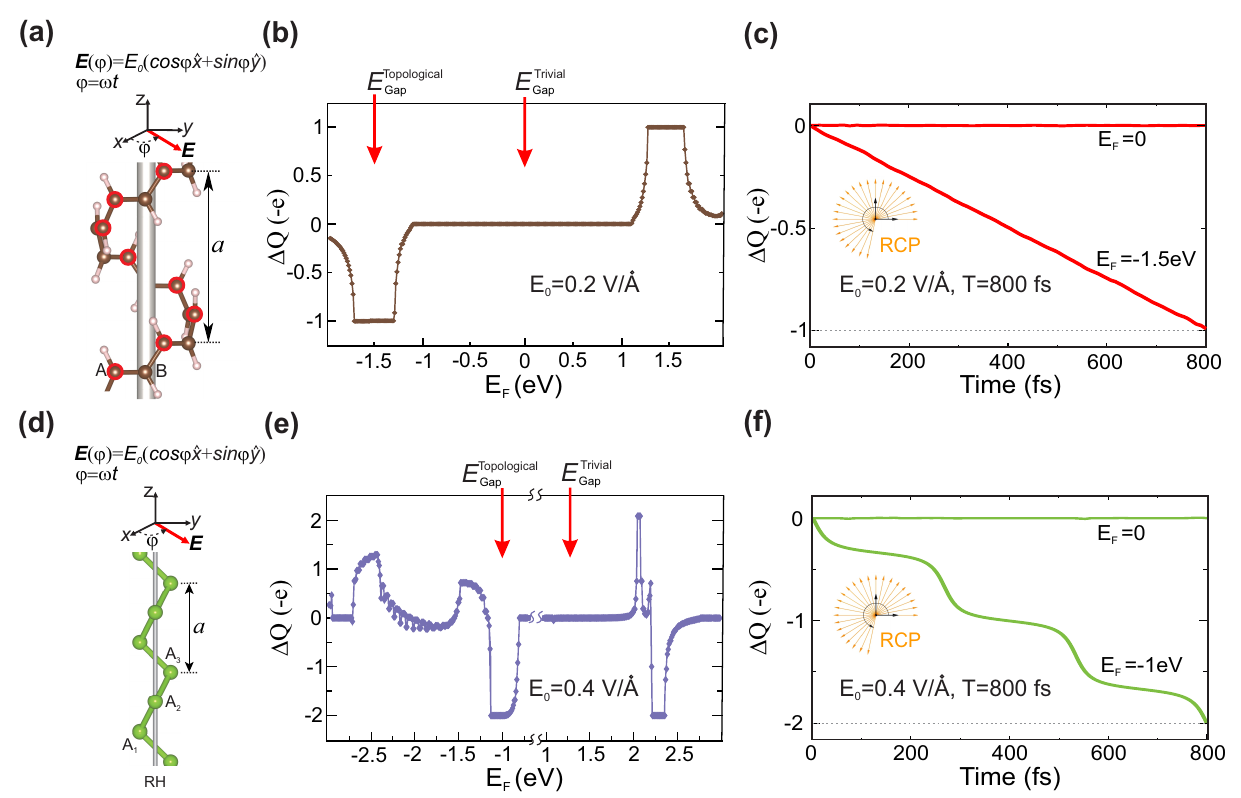} 
 \caption{{\bf $\bm{|}$ Numerical simulation of Thouless charge pumping in  chiral wires.}
  \textbf{a,d}, Right-handed chiral structures of (\textbf{a}) C$_{12}$H$_{12}$ and (\textbf{d}) trigonal Se, together with the applied circularly polarized driving field acting on each system.
 \textbf{b}, The transported charge per pump cycle $\Delta Q\,(-e)$ in the chiral hydrocarbon  C$_{12}$H$_{12}$ as a function of the Fermi energy, driven by the rotating electric field $\bm{E}(\varphi(t))$ with amplitude $E_0 = 0.2~\mathrm{V\AA^{-1}}$ and period $T=800~\mathrm{fs}$. 
\textbf{c}, Real-time profile of the Thouless pumped charge over one period $T$ with the Fermi level located within the $E_{\mathrm{gap}}^{\mathrm{Topological}}$ of $E_F=-1.5~\mathrm{eV}$.
\textbf{e,f}, Same as panels \textbf{c,d}, but for the trigonal chiral wire. The Fermi level 
(\textbf{f}) is set to $E_F=-1~\mathrm{eV}$ within the topological gap and the driving field  amplitude
is $E_0 = 0.4~\mathrm{V\AA^{-1}}$.
}  
 \label{fig:fig3} 
\end{figure}
 Because the screw symmetry is never restored at any point during the driving cycle, these field-induced gaps remain open during the full period \(T = 2\pi/\omega\).  These persistent gaps allow topological features to emerge and enable  modulating bands to contribute to   Thouless charge pumping. \\
 To demonstrate the topological nature of these gaps, we drive the two systems, shown in Fig.~\ref{fig:fig2}, 
 with a right-handed circularly polarized (RCP) electric field (Fig.~\ref{fig:fig3}a,d)  and probe their real-time charge dynamics.  Details of the current evaluation and the calculation of the pumped charge $\Delta Q$ over a full driving cycle are provided in the \textit{Methods} section.
Figure~\ref{fig:fig3}b presents the transported charge $\Delta Q\,(-e)$, over one pumping cycle in C$_{12}$H$_{12}$ for various Fermi levels. 
When the Fermi level lies within the trivial gap ($E_{\mathrm{gap}}^{\mathrm{Trivial}}$ in Fig.~\ref{fig:fig2}c), the pumped charge remains zero. However, it is very remarkable that when the Fermi level is located inside the 
field induced gap ($E_{\mathrm{gap}}^{\mathrm{Topological}}$ in Fig.~\ref{fig:fig2}c), each cycle of the drive produces exactly one quantum of pumped charge. 
The detailed real-time profile of the pumped charge over a single cycle is shown in Fig.\ref{fig:fig3}c.\\
The charge pumping along the SOC trigonal chiral wire exhibits
analogous dependence on the Fermi level, as presented in Fig.\ref{fig:fig3}e. One-cycle real-time profile of the charge for the case of non-trivial pumping is shown in Fig.\ref{fig:fig3}e. In both systems, the driving period $T=800~\mathrm{fs}$ (corresponding to $\hbar\omega\simeq5.2~\mathrm{meV}$) remains much smaller than the instantaneous band gaps during the entire cycle, ensuring adiabatic evolution by suppressing interband transitions.\\
\noindent In fact, the chiral geometry and its coupling to the field \(\bm{E}(\varphi)\) provide an adiabatic modulation of the sliding potential at frequency \(\omega\), thereby introducing a synthetic dimension through the phase parameter \(\varphi\). Together with the crystal momentum \(k\), the phase \(\varphi\) defines a synthetic \((1\!+\!1)\)-D parameter space \(\mathcal{R}=(k,\varphi)\in\mathbb{T}^2\). The resulting Berry curvature two-form on this
even-dimensional torus constitutes the integer Chern number corresponding to the quantized Thouless pump. 
Because the Chern number defined on $(k,\varphi)$ is equivalent to the 1D winding number of the Bloch band in the chiral wire, the pumping mechanism here is intrinsically different from that of the Rice-Mele model. Whereas Rice-Mele requires two independent modulations to open a gap and execute a nontrivial cycle, the geometric winding of the chiral system already provides the necessary phase structure. Consequently, only a single control parameter $\varphi$ is needed to maintain the gap and realize a quantized Thouless pump.

\begin{figure}[t!]
    \centering
    \includegraphics[width=.99\textwidth]{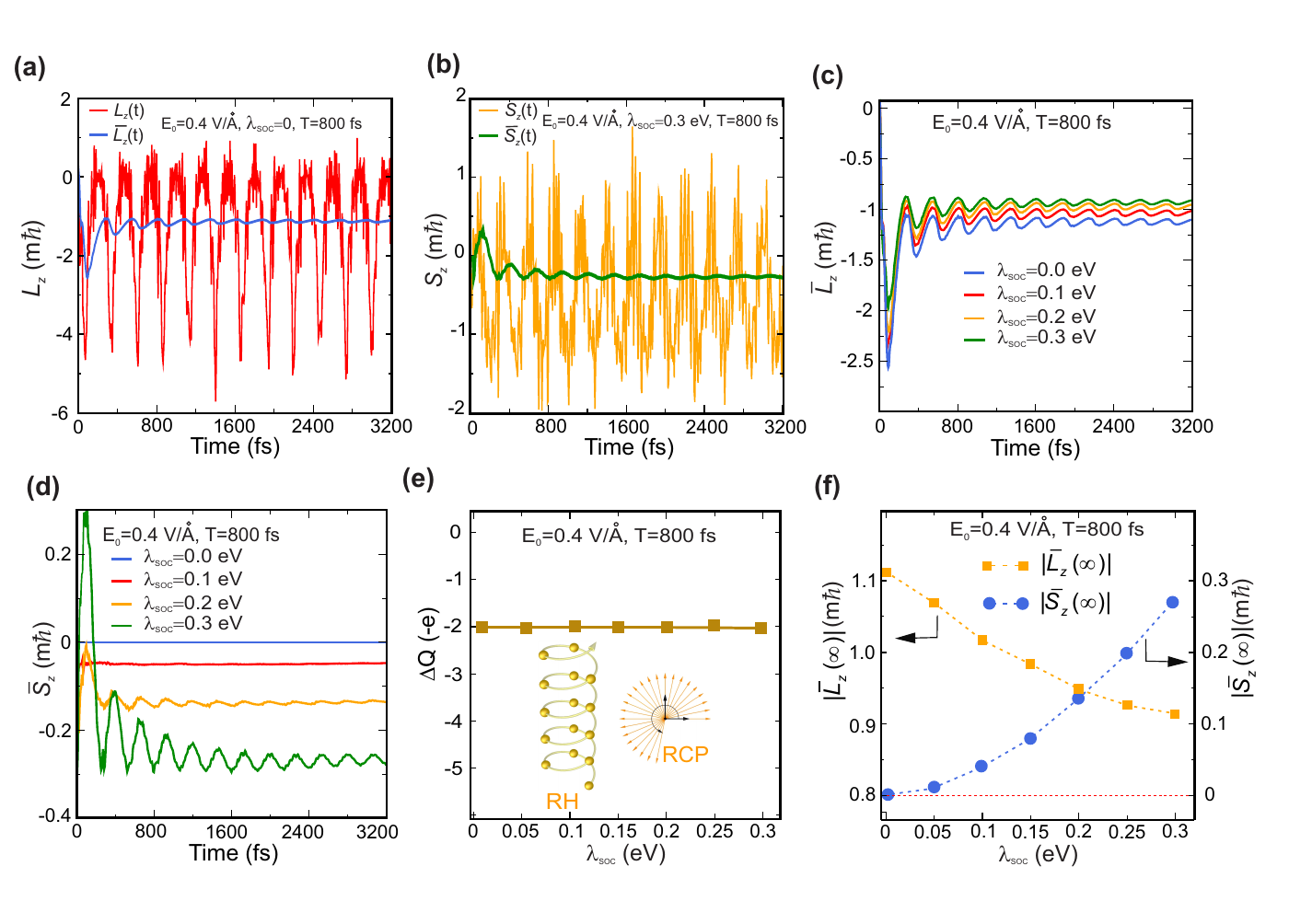} 
 \caption{{\bf $\bm{|}$  Orbital-spin polarization in a chiral wire during Thouless charge pumping.} \textbf{a}, Real-time profile of the OAM  $L_z(t)$ (red) and its time-averaged value $\bar{L}_z(t)$ (blue)  in a RH chiral wire undergoing topological pumping.  \textbf{b}, Corresponding SAM dynamics  $S_z(t)$ (yellow) and  its time-averaged value $\bar{S}_z(t)$ (green)
 for SOC strength $\lambda = 0.3~\mathrm{eV}$. 
\textbf{c},\textbf{d} Time-averaged orbital (\textbf{c}) and spin (\textbf{d}) polarizations as functions of SOC strength.
\textbf{e}, The pumped charge $\Delta Q$ over one cycle obtained with various SOC strength.
\textbf{f}, Time-averaged spin and orbital obtained after 4 cycles (denoted by  
 $|\bar{L}_z(\infty)|$ (solid squre) and $|\bar{S}_z(\infty)|$ (solid  ball).
}
 \label{fig:fig4} 
\end{figure}
\begin{figure}[t!]
    \centering
    \includegraphics[width=.95\textwidth]{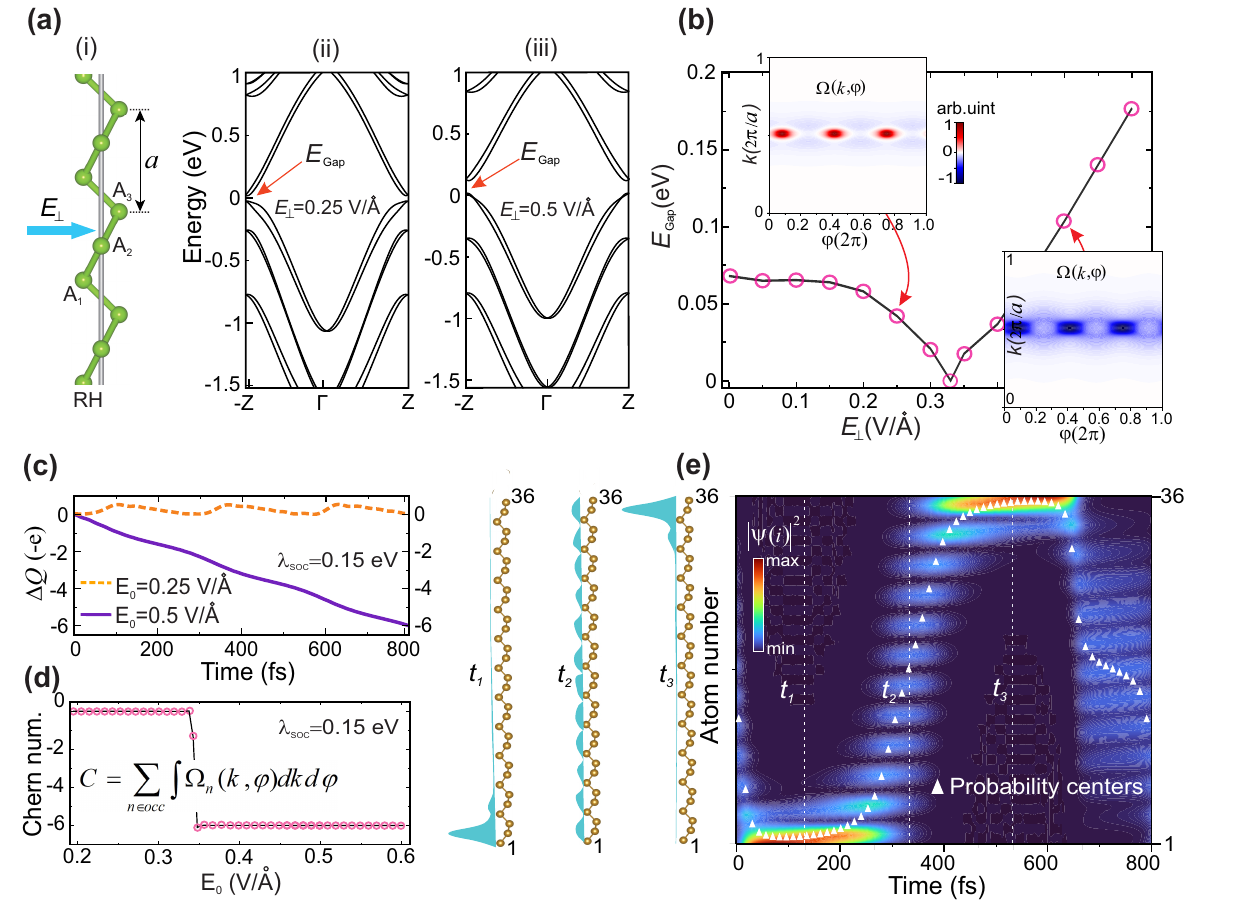} 
 \caption{{\bf $\bm{|}$  Topological phase transition in 1D a chiral wire.} \textbf{a}, (i) Schematic of a trigonal chiral wire subjected to a perpendicular electric field $E_\perp$.   Static band structures of the chiral wire calculated with parameters $V_{pp\sigma}^{12} = 1.5~\mathrm{eV}$, $V_{pp\pi}^{12} = -0.8~\mathrm{eV}$, and $\lambda = 0.15~\mathrm{eV}$, for field strengths (ii) $E_\perp = 0.25~\mathrm{V\AA^{-1}}$  and (iii) $E_\perp = 0.5~\mathrm{V\AA^{-1}}$. The band gaps are indicated by $E_{gap}$.
\textbf{b}, Evolution of the band gap as a function of the perpendicular field strength $E_\perp$. Insets show the Berry-curvature distributions in the synthetic $(k,\varphi)$ space for a rotating electric field with amplitudes $E_0 = 0.25~\mathrm{V\AA^{-1}}$ (top) and $E_0 = 0.5~\mathrm{V\AA^{-1}}$ (bottom).
\textbf{c}, Transported charge per pumping cycle obtained from real-time simulations along the chiral wire driven by a rotating electric field with amplitudes $E_0 = 0.25~\mathrm{V\AA^{-1}}$ (orange dashed curve) and $E_0 = 0.5~\mathrm{V\AA^{-1}}$ (violet curve).
\textbf{d}, The total Chern number in the synthetic $(k,\varphi)$ space as a function of the field amplitude.
\textbf{e}, Left: colormap of the site-resolved probability along a chain of $N=36$ atoms as a function of time, obtained by summing over occupied states with end-mode contributions. White triangles mark the centers of the probability distribution. Right: schematic snapshots of the electron probability amplitude along the chain at three representative times, $t_1$, $t_2$, and $t_3$.
}
	  
 \label{fig:fig5} 
\end{figure}

\noindent\textbf{  Orbital and spin angular momentum dynamics in a Thouless pump.}
Time-reversal symmetry enforces the antisymmetric orbital-momentum locking $L_z(n,k) = -L_z(n,-k)$ in  chiral structure. Orbital polarization effect (OPE) refers to the generation of OAM polarization in these materials when an applied bias field or driven current violates  this time-reversal protected constraint~\cite{liu2021chirality,yan2024structural}.
Although such an externally driven OAM has not yet been directly observed experimentally, this nonequilibrium orbital dynamics has recently been proposed as an underlying mechanism for CISS and nonreciprocal transport phenomena in chiral molecules and crystals ~\cite{liu2021chirality,adhikari2023interplay,yan2024structural}. Here, we extend the concept of OPE beyond metallic charge transport to the regime of topological Thouless charge pumping, specifically investigating how OAM accompanies and responds to the quantized topological charge pump. We begin by examining the Heisenberg equation of motion for spin and orbitals.
As described in Supplementary Note 1 (Eq.~S11), the effective Hamiltonian in the synthetic (1+1)-dimensional 
can be derived from Eq.~(\ref{eq:eq6_1}) in the following form:
\begin{equation}
\hat{H}(k,\varphi)
= \hat{I}_{2\times2}^{(\mathrm{spin})}\!\otimes\!\hat{T}(k)
+ \lambda_{\mathrm{SOC}}\,\hat{\bm{\sigma}}\!\cdot\!\hat{\mathbf{L}}
- e\,\mathbf{E}(\varphi)\!\cdot\!\hat{\mathbf{r}}.
\label{eq:H_compact}
\end{equation}
Detailed derivation of the Heisenberg equation of motions are summarized in Supplementary Note 6, which can be summarized as
\begin{align}
\frac{d\hat{\mathbf{L}}}{dt}
&=\frac{i}{\hbar}[\hat{H}(k,\varphi),\hat{L}]
= e\,\hat{\mathbf{r}}\!\times\!\mathbf{E}(\varphi)
+ \lambda_{\mathrm{SOC}}\,\hat{\mathbf{L}}\!\times\!\hat{\bm{\sigma}}
+ \frac{i}{\hbar}\,[\hat{I}_{2\times2}^{(\mathrm{spin})}\!\otimes\!\hat{T}(k),\hat{\mathbf{L}}],
\label{eq:dLdt_SOC} \\[4pt]
\frac{d\hat{\mathbf{S}}}{dt}
&=\frac{i}{\hbar}[\hat{H}(k,\varphi),\hat{S}]= \lambda_{\mathrm{SOC}}\,\hat{\bm{\sigma}}\!\times\!\hat{\mathbf{L}}.
\label{eq:dSdt_SOC}
\end{align}
Important lessons can be derived from these equations. The longitudinal component of orbital
\( L_z(t) \) is driven by the OPE torque \( \hat{\mathbf{r}} \times \mathbf{E}(\varphi) \) with the in-plane rotating driving field
even when the SOC is absent (\( \lambda_{\mathrm{SOC}} = 0 \)). The rate of spin generation is linearly proportional to the SOC-induced torque
\( \hat{\boldsymbol{\sigma}} \times \hat{\mathbf{L}} \), which is adversely affected by the rate of orbital generation
(\( \hat{\mathbf{L}} \times \hat{\boldsymbol{\sigma}} = - \hat{\boldsymbol{\sigma}} \times \hat{\mathbf{L}} \)).
By explicitly calculating the time-dependent Schrödinger equation, we traced the real-time evolution of the OAM and SAM during topological charge pumping in a trigonal chiral wire, as shown in Fig.~\ref{fig:fig4}.  Figure~\ref{fig:fig4}a displays the OAM dynamics in the system without SOC with the Fermi level sitting in the topological gap (see Fig.~\ref{fig:fig3}a).
As seen, the instantaneous polarization $L_z(t)$ (red) shows an oscillatory profile, and its time-averaged value (blue), $\bar{L}_z =\frac{1}{t - t_0}\!\int_{t_0}^{t} L_z(t')\,dt'$, gradually develops and stabilizes over successive pumping cycles. This behavior provides direct evidence of OPE induced by the  Thouless charge pumping in chiral wires. 
When SOC is included, part of the OAM generated by the OPE torque is converted into spin.
As depicted in Fig.~\ref{fig:fig4}b, the SAM polarization $S_z(t)$ (yellow) exhibits a similar oscillatory profile as the orbital dynamics, and its time-averaged value (green), $\bar{S}_z =\frac{1}{t - t_0}\!\int_{t_0}^{t} S_z(t')\,dt'$, reveals the emergence of a net spin polarization over several  pumping cycles.
To elucidate the effect of SOC in the interconversion behavior between OAM and SAM, we simulated the time-evolution with various strengths of SOC: the results of the time-averaged spin and orbital are presented in Figs.~\ref{fig:fig4}c and d. As expected from the term SOC-induced torque ($\,\hat{\mathbf{L}}\times\hat{\bm{\sigma}}$=-$\hat{\bm{\sigma}}\times\hat{\mathbf{L}}$ in Eq.~\ref{eq:dLdt_SOC})
$\bar{S}_z$ monotonically increases with the strength of SOC, whereas the $\bar{L}_z$ decreases. The time-averaged values after 4 cycles (3200~fs) are denoted by $|\bar{L}_z(\infty)|$ (solid square) and $|\bar{S}_z(\infty)|$ (solid ball) and presented in Fig.~\ref{fig:fig4}f with various SOC strengths.\\
Remarkably, the amount of pumped charge over one cycle remains as one quantum, irrespective of the SOC strength, which obviously proves the topological nature of the charge pumping. These results demonstrate that the orbital dynamics in the chiral Thouless pump are governed predominantly by the OPE torque 
$e\,\hat{\mathbf{r}}\times\mathbf{E}(\varphi)$. Together with the intrinsic orbital-momentum locking term, this reveals a purely geometric consequence of the structural chirality. By contrast, the spin response arises solely through SOC, which converts the OAM generated by the OPE mechanism into SAM. Consequently, our findings suggest that a topologically quantized charge pump in insulating chiral wires can generate spin polarization via this orbital-to-spin conversion pathway, suggesting a distinct insight on the CISS effect.\\
\noindent\textbf{Thouless topological phase transition in a chiral wire.}
Above in Fig.~\ref{fig:fig2} and Fig.~\ref{fig:fig3}, we demonstrated that the perpendicular bias field ($E_\perp$) opens a topological gap ($E_{\mathrm{Gap}}^{\mathrm{topological}}$) and induces the Thouless charge pumping, whose topological nontriviality depends on the selection of the Fermi level.
We now address the  question of  whether applying a perpendicular electric field to a chiral wire can induce band inversion and thereby  tune the  topological character of its originally trivial gap ($E_{\mathrm{Gap}}^{\mathrm{trivial}}$ in Figs.~\ref{fig:fig2}c and f)? To mimic the typical reduced band gap in a bundle of 1D trigonal wire~\cite {kramer2020tellurium},  we adopt a refined parameter set
with hopping amplitudes $V_{pp\sigma}^{12} = 1.5$~eV and $V_{pp\pi}^{12} = -0.8$~eV (see Supplementary Note 1). This choice preserves the essential features of the electronic structure in a trigonal wire while producing a narrower gap. 
Figure~\ref{fig:fig5}a presents the static band structure of the trigonal chiral wire under a perpendicular electric field (see panel i), for two representative field strengths, $E_\perp = 0.25~\mathrm{V\AA^{-1}}$ (panel ii) and $E_\perp = 0.5~\mathrm{V\AA^{-1}}$ (panel iii).
 The corresponding field-dependent evolution of the gap is summarized in Fig.~\ref{fig:fig5}b. Notably, the originally trivial band gap ($E_{\mathrm{Gap}}^{\mathrm{trivial}}$) closes at $E_\perp = 0.33~\mathrm{V\AA^{-1}}$, and it reopens upon further increasing the field. This behavior indicates a field-driven band inversion in the electronic spectrum.  We now show that this reopened gap becomes topological.
Figure~\ref{fig:fig5}c displays the real-time dynamics of the transported charge per pumping cycle. At $E_0 = 0.25~\mathrm{V\AA^{-1}}$, the total pumped charge  (orange curve) vanishes, arising from the topologically trivial nature of the band gap. In contrast, at field amplitude $E_0 = 0.5~\mathrm{V\AA^{-1}}$, the contributions from the Kramers doublets no longer cancel, yielding a finite net pumped charge (violet curve) of $\Delta Q = -6$(-e).
To further confirm the topological character, we calculate the Berry curvature $\Omega(k,\varphi)$ (see Methods) of the static band structure defined in the $(1+1)$-dimensional Brillouin zone ($0 \le k \le 2\pi/a$, $0 \le \varphi \le 2\pi$). As shown in the inset of Fig.~\ref{fig:fig5}b, the Berry curvatures corresponding to the trivial and topological responses in Fig.~\ref{fig:fig5}c exhibit qualitatively distinct behaviors. In the trivial phase, the Berry curvature develops sharp positive peaks (red) accompanied by a diffuse negative background (blue) whose contributions give zero integration
over the full Brillouin zone. By contrast, under the stronger drive, the Berry-curvature profile is reshaped, indicative of an inverted band ordering and resulting in a finite quantized value. 
Figure~\ref{fig:fig5}d summarizes the total Chern number as a function of field strength, clearly revealing the associated Thouless topological phase transition.
\noindent To further elucidate Thouless charge pumping, we compute the site-resolved probability amplitudes in a finite chiral chain. Let \( |\Psi_n(t)\rangle \) denote the instantaneous (time-evolved) wave function of the \(n\)-th state in a finite system comprising \(N\) atoms. The site resolved probability for the $n$-th state at the site $i$ is calculated  as
\begin{align}
 p_{n}^{i}(t)
&     = \sum_{\mu \sigma } 
\langle \Psi_n(t) \,|\,i,\mu,\sigma\ \rangle
\langle i,\mu,\sigma\,|\, \Psi_n(t) \rangle \notag \\
& = \sum_{\mu }
\left( \left| \phi_{n,\mu\uparrow}^{i}(t) \right|^2 
+ \left| \phi_{n,\mu\downarrow}^{i}(t) \right|^2 \right),
\label{eq:eq2}
\end{align}
  where $|\,i,\mu,\sigma\ \rangle$  denotes the local orbital $\mu\in {\{p_x, p_y, p_z}\}$ with spin $\sigma$ centered at the site $i$.
  Figure \ref{fig:fig5}e (left) shows the colormap of the site-resolved probability along a chain of 
$N=36$ atoms as a function of time, obtained by summing over the occupied states that contain end-mode contributions. We define this summed probability as
$P_i(t) = \sum\limits_{n\in{em}}f_n^{em}p_{n}^{i}(t)$, where $f_n^{em}$ denotes the occupation factor of the end-mode states.
From this site-resolved probability, we obtain the time-dependent center of the wave-packet (white triangles in Fig.~\ref{fig:fig5}e) along the chain as  
$z_c(t) = \frac{1}{N_{\mathcal{B}}}\sum\limits_{i=1}^{N} z^i \, P_i(t),$
where $z^i$ is the position of the site $i$ along the chain, and $N_{\mathcal{B}}$ is the number of occupied states with end-mode components.  The figure corresponds to the parameter set of the topological phase. The evolution of the site-resolved probability and the corresponding wave-packet center clearly reveal the transport characteristic of the pumping process. In the right panel of  the figure, the schematic snapshots at $t_1$, $t_2$, and $t_3$ highlight representative configurations of the probability profile during the cycle.
\section{Discussion}\label{Discussion}
We demonstrated that a one-dimensional chiral system provides a real-material
realization of Thouless pumping, in which quantized topological charge transport is
accompanied by the transfer of orbital and spin angular momentum. The key ingredient
is the Dirac-cone-type degeneracy protected by the material’s screw-rotation symmetry,
together with its controlled breaking by an external perpendicular electric field. Using the
time-evolving Schrödinger equation of a multi-orbital TB Hamiltonian, we
obtained real-time profiles of the charge current, orbital angular momentum, and spin. In
parallel, we calculated the Berry curvature and topological invariants defined on the
even-dimensional (1+1)-dimensional synthetic space of the insulator. The torque
mechanism, originating from the synergistic combination of geometrical chirality and
field polarization, predominantly drives the orbital response, which is subsequently
converted into spin angular momentum through spin-orbit coupling. We discussed that
this orbital-spin interconversion provides a bulk topological pathway toward the
chirality-induced spin selectivity (CISS) effect, rooted in the Berry-phase physics of
topological charge transport of the insulator. We further suggest that this system offers a
salient example in which bulk topological numbers may be identified through directly
measurable physical observables-such as orbital or spin angular momentum—without
relying on bulk-edge correspondence.

\section{Methods}\label{Methods}
\textbf{Electronic structure calculation.} First-principles calculations were performed within density functional theory using the Quantum ESPRESSO package. The exchange-correlation functional was described by the Perdew-Burke-Ernzerhof functional. Electron-ion interactions were treated using a norm-conserving pseudopotential. The energy cutoff for the plane-wave basis was set to 60 Ry for wavefunctions and 240 Ry for charge density. The simulation cell for the chiral hydrocarbon structure contains 24 atoms, and Brillouin-zone sampling was carried out using a 1×1×31 Monkhorst-Pack k-point mesh for the consideration of the isolated one-dimensional chiral wire. A vacuum spacing of 9Å was introduced in the in-plane direction to simulate the effect of the isolated 1D wire. Additionally, for the trigonal chiral Se wire, we used 1×1×61 k-mesh and the same vacuum spacing for the isolation of the chain. \\
\textbf{Real-time calculations.}
We investigate the charge, spin, and orbital dynamics of the chiral wire by performing the real-time time-dependent TB Schrödinger equation 
\begin{equation}
i\hbar \frac{\partial}{\partial t} |\psi_{n,k}(t)\rangle = \hat{H}[\varphi(t)] |\psi_{n,k}(t)\rangle,
\end{equation}
where $|\psi_{n,k}(t)\rangle$ is the time-evolved Bloch state of band $n$ and momentum $k$.
The time-dependent Hamiltonian $\hat{H}[\varphi(t)]$ is given by Eq.~(S11) and includes the coupling to a circularly polarized electric field through the phase parameter $\varphi(t)$.
The time profile of the electric current density along the wire is calculated as
\begin{equation}
J(t) = -\frac{e}{a}\sum_{n,k} f_{n,k}
\langle \psi_{n,k}(t) | \hat{\mathbf{v}} | \psi_{n,k}(t) \rangle,
\end{equation}
where $a$ is the lattice constant, $f_{n,k}$ is the initial band occupation, and 
$\hat{\mathbf{v}} = \frac{1}{\hbar}\frac{\partial \hat{H}}{\partial k}$ 
denotes the band velocity operator. 
The total charge pumped during one full cycle of the drive is obtained through
\begin{equation}
\Delta Q = \int_0^T J(t)\,dt.
\end{equation}
\\
For a 1D chiral wire with $N$-fold screw symmetry, we employ the spin-major basis
$
\mathcal{B} = \{ \uparrow, \downarrow \} \otimes\{ A_1, A_2, \ldots, A_N \}\otimes \{ p_x, p_y, p_z \},  
$
where $A_i$ labels the atomic sites within the unit cell. In this basis, the total spin operator for each component $\alpha = x, y, z$ acts only in the spin subspace and is defined as
\begin{equation}
\hat{S}_\alpha = \frac{\hbar}{2}\,\hat{\sigma}_\alpha\otimes \hat{I}^{(\mathrm{atom})}_{N\times N} \otimes \hat{I}^{(\mathrm{orb})}_{3\times3},
\end{equation}
where $\hat{\sigma}_\alpha$ are the Pauli matrices, and $\hat{I}^{(\mathrm{atom})}$ and $\hat{I}^{(\mathrm{orb})}$  are identity operators in the atom and orbital  subspaces, respectively. 
The SAM dynamics for component $\alpha$ is computed by
\begin{equation}
S_\alpha(t) = \sum_{n,k} f_{n,k} 
\langle \psi_{n,k}(t) | \hat{S}_\alpha | \psi_{n,k}(t) \rangle.
\end{equation} 
%
\noindent To analyze the orbital response, we evaluate the dynamics of the OAM using the atom-centered approximation. The $\alpha$-th component of the total OAM operator can be expressed as
\begin{equation}
\hat{L}_{\alpha}
=
\hat{I}_{2\times 2}^{(\mathrm{spin})}
\otimes
\Bigl[
  \hat{L}_{\alpha}^{(A_1)}
  \oplus
  \hat{L}_{\alpha}^{(A_2)}
  \oplus
  \cdots
  \oplus
  \hat{L}_{\alpha}^{(A_N)}
\Bigr],
\end{equation}
where $\hat{I}_{2\times 2}^{(\mathrm{spin})}$ denotes the identity operator in spin space, and $\hat{L}_{\alpha}^{(A_i)}$ represents the local OAM operator at atomic site $A_i$. The explicit matrix representation of $\hat{L}_{\alpha}$ in the $\{p_x,p_y,p_z\}$ orbital basis is given in Eqs.~(S6).
 The time-dependent OAM expectation value is then obtained as  
\begin{equation}
L_\alpha(t) = \sum_{k} \sum_n f_{n,k} 
\langle \psi_{n,k}(t) | \hat{L}_\alpha | \psi_{n,k}(t) \rangle.
\end{equation}
\noindent\textbf{Chern number in the synthetic space.}
To characterize topological features in our driven 1D chiral system, we compute the Chern number defined in the synthetic (1+1)-dimensional parameter space \((k, \varphi)\), where \(k\) denotes the crystal momentum along the physical dimension and \(\varphi\) is the phase of the applied circularly polarized electric field, treated as a synthetic dimension.
The Berry curvature associated with the \(n\)-th energy band can be expressed as\cite{xiao2010berry,wang2006ab,vanderbilt2018berry}
\begin{equation}
\Omega_n(k,\varphi) = -2\,\mathrm{Im} \sum_{m \neq n} \frac{\langle u_{nk} | \partial_k \hat{H} | u_{mk} \rangle \langle u_{mk} | \partial_\varphi \hat{H} | u_{nk} \rangle}{(\varepsilon_{mk} - \varepsilon_{nk})^2}.
\end{equation}
Here, \(\varepsilon_{nk}\) denotes the corresponding instantaneous band energy, while \(\partial_k \hat{H}\) and \(\partial_\varphi \hat{H}\) represent the derivatives of the Hamiltonian with respect to \(k\) and \(\varphi\), respectively. 
The total Berry curvature is given by \cite{xiao2010berry,wang2006ab,vanderbilt2018berry}
\begin{equation}
\Omega(k,\varphi) = \sum_{n} f_{n,k}\,\Omega_n(k,\varphi),
\end{equation}
where $f_{n,k}$ denotes the occupation factor of band $n$. The associated Chern number is obtained by integrating the Berry curvature over the synthetic $(k,\varphi)$ Brillouin zone,
\begin{equation}
C= \int_{0}^{2\pi} d\varphi \int_{\mathrm{BZ}} \Omega(k,\varphi)\, dk.
\end{equation}
The resulting change in polarization over a full driving cycle, corresponding to the net pumped charge, is then given by \cite{vanderbilt2018berry}
\begin{equation}
\Delta Q=\Delta P = -e\, C.
\end{equation}
\section{Data availability}\label{Data availability}
The main data supporting the findings of this study are available within the article and
its Supplementary Information files. All the raw data generated in this study are available
from the corresponding authors upon reasonable request.
\section{Code availability}\label{Code availability}
All the data analysis codes related to this study are available from the corresponding
authors upon reasonable request. Codes used to produce the data will be provided under
guarantee of acknowledgements/appropriate citation of this work, and a scientifically
sound reason for request.

\bibliography{main_refs}

@article{laughlin1981quantized,
  title={Quantized Hall conductivity in two dimensions},
  author={Laughlin, Robert B},
  journal={Physical Review B},
  volume={23},
  number={10},
  pages={5632},
  year={1981},
  publisher={APS}
}

@article{thouless1983quantization,
  title={Quantization of particle transport},
  author={Thouless, DJ},
  journal={Physical Review B},
  volume={27},
  number={10},
  pages={6083},
  year={1983},
  publisher={APS}
}

@article{wang2013topological,
  title={Topological charge pumping in a one-dimensional optical lattice},
  author={Wang, Lei and Troyer, Matthias and Dai, Xi},
  journal={Physical review letters},
  volume={111},
  number={2},
  pages={026802},
  year={2013},
  publisher={APS}
}

@article{nakajima2016topological,
  title={Topological Thouless pumping of ultracold fermions},
  author={Nakajima, Shuta and Tomita, Takafumi and Taie, Shintaro and Ichinose, Tomohiro and Ozawa, Hideki and Wang, Lei and Troyer, Matthias and Takahashi, Yoshiro},
  journal={Nature Physics},
  volume={12},
  number={4},
  pages={296--300},
  year={2016},
  publisher={Nature Publishing Group UK London}
}

@article{lohse2016thouless,
  title={A Thouless quantum pump with ultracold bosonic atoms in an optical superlattice},
  author={Lohse, Michael and Schweizer, Christian and Zilberberg, Oded and Aidelsburger, Monika and Bloch, Immanuel},
  journal={Nature Physics},
  volume={12},
  number={4},
  pages={350--354},
  year={2016},
  publisher={Nature Publishing Group UK London}
}

@article{nakajima2021competition,
  title={Competition and interplay between topology and quasi-periodic disorder in Thouless pumping of ultracold atoms},
  author={Nakajima, Shuta and Takei, Nobuyuki and Sakuma, Keita and Kuno, Yoshihito and Marra, Pasquale and Takahashi, Yoshiro},
  journal={Nature Physics},
  volume={17},
  number={7},
  pages={844--849},
  year={2021},
  publisher={Nature Publishing Group UK London}
}

@article{sun2022non,
  title={Non-Abelian Thouless pumping in photonic waveguides},
  author={Sun, Yi-Ke and Zhang, Xu-Lin and Yu, Feng and Tian, Zhen-Nan and Chen, Qi-Dai and Sun, Hong-Bo},
  journal={Nature Physics},
  volume={18},
  number={9},
  pages={1080--1085},
  year={2022},
  publisher={Nature Publishing Group UK London}
}

@article{grinberg2020robust,
  title={Robust temporal pumping in a magneto-mechanical topological insulator},
  author={Grinberg, Inbar Hotzen and Lin, Mao and Harris, Cameron and Benalcazar, Wladimir A and Peterson, Christopher W and Hughes, Taylor L and Bahl, Gaurav},
  journal={Nature communications},
  volume={11},
  number={1},
  pages={974},
  year={2020},
  publisher={Nature Publishing Group UK London}
}

@article{chen2021topological,
  title={Topological pumping in acoustic waveguide arrays with hopping modulation},
  author={Chen, Zhaoxian and Chen, Zeguo and Li, Zhengwei and Liang, Bin and Ma, Guancong and Lu, Yanqing and Cheng, Jianchun},
  journal={New Journal of Physics},
  volume={24},
  number={1},
  pages={013004},
  year={2021},
  publisher={IOP Publishing}
}

@article{song2024fast,
  title={Fast topological pumps via quantum metric engineering on photonic chips},
  author={Song, Wange and You, Oubo and Sun, Jiacheng and Wu, Shengjie and Chen, Chen and Huang, Chunyu and Qiu, Kai and Zhu, Shining and Zhang, Shuang and Li, Tao},
  journal={Science Advances},
  volume={10},
  number={30},
  pages={eadn5028},
  year={2024},
  publisher={American Association for the Advancement of Science}
}

@article{haug2019topological,
  title={Topological pumping in Aharonov--Bohm rings},
  author={Haug, Tobias and Dumke, Rainer and Kwek, Leong-Chuan and Amico, Luigi},
  journal={Communications Physics},
  volume={2},
  number={1},
  pages={127},
  year={2019},
  publisher={Nature Publishing Group UK London}
}

@article{wang2022two,
  title={Two-dimensional Thouless pumping of light in photonic moir{\'e} lattices},
  author={Wang, Peng and Fu, Qidong and Peng, Ruihan and Kartashov, Yaroslav V and Torner, Lluis and Konotop, Vladimir V and Ye, Fangwei},
  journal={Nature communications},
  volume={13},
  number={1},
  pages={6738},
  year={2022},
  publisher={Nature Publishing Group UK London}
}

@article{rice1982elementary,
  title={Elementary excitations of a linearly conjugated diatomic polymer},
  author={Rice, MJ and Mele, EJ},
  journal={Physical Review Letters},
  volume={49},
  number={19},
  pages={1455},
  year={1982},
  publisher={APS}
}

@article{guo2017topological,
  title={Topological states and quantized current in helical organic molecules},
  author={Guo, Ai-Min and Sun, Qing-Feng},
  journal={Physical Review B},
  volume={95},
  number={15},
  pages={155411},
  year={2017},
  publisher={APS}
}

@article{guo2020topological,
  title={Topological phase transitions of Thouless charge pumping realized in helical organic molecules with long-range hopping},
  author={Guo, Ai-Min and Hu, Pei-Jia and Gao, Xiao-Hui and Fang, Tie-Feng and Sun, Qing-Feng},
  journal={Physical Review B},
  volume={102},
  number={15},
  pages={155402},
  year={2020},
  publisher={APS}
}

@article{evers2022theory,
  title={Theory of chirality induced spin selectivity: Progress and challenges},
  author={Evers, Ferdinand and Aharony, Amnon and Bar-Gill, Nir and Entin-Wohlman, Ora and Hedeg{\aa}rd, Per and Hod, Oded and Jelinek, Pavel and Kamieniarz, Grzegorz and Lemeshko, Mikhail and Michaeli, Karen and others},
  journal={Advanced Materials},
  volume={34},
  number={13},
  pages={2106629},
  year={2022},
  publisher={Wiley Online Library}
}

@article{adhikari2023interplay,
  title={Interplay of structural chirality, electron spin and topological orbital in chiral molecular spin valves},
  author={Adhikari, Yuwaraj and Liu, Tianhan and Wang, Hailong and Hua, Zhenqi and Liu, Haoyang and Lochner, Eric and Schlottmann, Pedro and Yan, Binghai and Zhao, Jianhua and Xiong, Peng},
  journal={Nature communications},
  volume={14},
  number={1},
  pages={5163},
  year={2023},
  publisher={Nature Publishing Group UK London}
}

@article{naaman2019chiral,
  title={Chiral molecules and the electron spin},
  author={Naaman, Ron and Paltiel, Yossi and Waldeck, David H},
  journal={Nature Reviews Chemistry},
  volume={3},
  number={4},
  pages={250--260},
  year={2019},
  publisher={Nature Publishing Group UK London}
}

@article{bloom2024chiral,
  title={Chiral induced spin selectivity},
  author={Bloom, Brian P and Paltiel, Yossi and Naaman, Ron and Waldeck, David H},
  journal={Chemical Reviews},
  volume={124},
  number={4},
  pages={1950--1991},
  year={2024},
  publisher={ACS Publications}
}

@article{liu2021chirality,
  title={Chirality-driven topological electronic structure of DNA-like materials},
  author={Liu, Yizhou and Xiao, Jiewen and Koo, Jahyun and Yan, Binghai},
  journal={Nature Materials},
  volume={20},
  number={5},
  pages={638--644},
  year={2021},
  publisher={Nature Publishing Group UK London}
}

@article{abendroth2019spin,
  title={Spin selectivity in photoinduced charge-transfer mediated by chiral molecules},
  author={Abendroth, John M and Stemer, Dominik M and Bloom, Brian P and Roy, Partha and Naaman, Ron and Waldeck, David H and Weiss, Paul S and Mondal, Prakash Chandra},
  journal={ACS nano},
  volume={13},
  number={5},
  pages={4928--4946},
  year={2019},
  publisher={ACS Publications}
}

@article{guo2014spin,
  title={Spin-dependent electron transport in protein-like single-helical molecules},
  author={Guo, Ai-Min and Sun, Qing-Feng},
  journal={Proceedings of the National Academy of Sciences},
  volume={111},
  number={32},
  pages={11658--11662},
  year={2014},
  publisher={National Academy of Sciences}
}

@article{kim2023optoelectronic,
  title={Optoelectronic manifestation of orbital angular momentum driven by chiral hopping in helical Se chains},
  author={Kim, Bumseop and Shin, Dongbin and Namgung, Seon and Park, Noejung and Kim, Kyoung-Whan and Kim, Jeongwoo},
  journal={ACS nano},
  volume={17},
  number={19},
  pages={18873--18882},
  year={2023},
  publisher={ACS Publications}
}

@article{yan2024structural,
  title={Structural chirality and electronic chirality in quantum materials},
  author={Yan, Binghai},
  journal={Annual Review of Materials Research},
  volume={54},
  year={2024},
  publisher={Annual Reviews}
}

@article{lu2016geometrical,
  title={Geometrical pumping with a Bose-Einstein condensate},
  author={Lu, H-I and Schemmer, Max and Aycock, Lauren M and Genkina, Dina and Sugawa, Seiji and Spielman, Ian B},
  journal={Physical review letters},
  volume={116},
  number={20},
  pages={200402},
  year={2016},
  publisher={APS}
}

@article{walter2023quantization,
  title={Quantization and its breakdown in a Hubbard--Thouless pump},
  author={Walter, Anne-Sophie and Zhu, Zijie and G{\"a}chter, Marius and Minguzzi, Joaqu{\'\i}n and Roschinski, Stephan and Sandholzer, Kilian and Viebahn, Konrad and Esslinger, Tilman},
  journal={Nature Physics},
  volume={19},
  number={10},
  pages={1471--1475},
  year={2023},
  publisher={Nature Publishing Group UK London}
}

@article{kramer2020tellurium,
  title={Tellurium as a successor of silicon for extremely scaled nanowires: a first-principles study},
  author={Kramer, Aaron and Van de Put, Maarten L and Hinkle, Christopher L and Vandenberghe, William G},
  journal={npj 2D Materials and Applications},
  volume={4},
  number={1},
  pages={10},
  year={2020},
  publisher={Nature Publishing Group UK London}
}

@article{xiao2010berry,
  title={Berry phase effects on electronic properties},
  author={Xiao, Di and Chang, Ming-Che and Niu, Qian},
  journal={Reviews of modern physics},
  volume={82},
  number={3},
  pages={1959--2007},
  year={2010},
  publisher={APS}
}

@article{wang2006ab,
  title={Ab initio calculation of the anomalous Hall conductivity by Wannier interpolation},
  author={Wang, Xinjie and Yates, Jonathan R and Souza, Ivo and Vanderbilt, David},
  journal={Physical Review B—Condensed Matter and Materials Physics},
  volume={74},
  number={19},
  pages={195118},
  year={2006},
  publisher={APS}
}

@book{vanderbilt2018berry,
  title={Berry phases in electronic structure theory: electric polarization, orbital magnetization and topological insulators},
  author={Vanderbilt, David},
  year={2018},
  publisher={Cambridge University Press}
}

@article{starkloff1978electronic,
  title={The electronic structure of trigonal Se and Te for pressure near the phase transition points},
  author={Starkloff, Th and Joannopoulos, JD},
  journal={The Journal of Chemical Physics},
  volume={68},
  number={2},
  pages={579--584},
  year={1978},
  publisher={American Institute of Physics}
}

@article{qin2020raman,
  title={Raman response and transport properties of tellurium atomic chains encapsulated in nanotubes},
  author={Qin, Jing-Kai and Liao, Pai-Ying and Si, Mengwei and Gao, Shiyuan and Qiu, Gang and Jian, Jie and Wang, Qingxiao and Zhang, Si-Qi and Huang, Shouyuan and Charnas, Adam and others},
  journal={Nature electronics},
  volume={3},
  number={3},
  pages={141--147},
  year={2020},
  publisher={Nature Publishing Group UK London}
}

@article{qiu2022resurrection,
  title={The resurrection of tellurium as an elemental two-dimensional semiconductor},
  author={Qiu, Gang and Charnas, Adam and Niu, Chang and Wang, Yixiu and Wu, Wenzhuo and Ye, Peide D},
  journal={npj 2D Materials and Applications},
  volume={6},
  number={1},
  pages={17},
  year={2022},
  publisher={Nature Publishing Group UK London}
}

@article{nakayama2024observation,
  title={Observation of edge states derived from topological helix chains},
  author={Nakayama, K and Tokuyama, A and Yamauchi, K and Moriya, A and Kato, T and Sugawara, K and Souma, S and Kitamura, M and Horiba, K and Kumigashira, H and others},
  journal={Nature},
  volume={631},
  number={8019},
  pages={54--59},
  year={2024},
  publisher={Nature Publishing Group UK London}
}

@article{xie2011spin,
  title={Spin specific electron conduction through DNA oligomers},
  author={Xie, Zouti and Markus, Tal Z and Cohen, Sidney R and Vager, Zeev and Gutierrez, Rafael and Naaman, Ron},
  journal={Nano letters},
  volume={11},
  number={11},
  pages={4652--4655},
  year={2011},
  publisher={ACS Publications}
}


\section*{Supplementary material}

# Supplementary Notes For:

# Spin and Orbital Angular Momentum Transfer in Thouless Topological Charge Pumping

Esmaeil Taghizadeh Sisakht,[†] Uiseok Jeong,[†] Xiao jiang,[†] Jinseok Oh,[†] Yizhou Liu,[‡] Binghai Yan,[*,¶] and Noejung Park[*,†]

[†]Department of Physics, Ulsan National Institute of Science and Technology, Ulsan, 689-798, Korea

[‡]Center for Phononics and Thermal Energy Science, China-EU Joint Lab on Nanophononics, School of Physics Science and Engineering, Tongji University, 200092 Shanghai, China

[¶]Department of Physics, Pennsylvania State University, 201 Old Main, University Park, Pennsylvania, 16802, USA

E-mail: binghai.yan@psu.edu; noejung@unist.ac.kr

## Supplementary Note 1: Multi-orbital tight-binding model for a chiral wire

In the main text, we introduced the real space tight-binding (TB) Hamiltonian (Eq.(1)) for a one-dimensional (1D) chiral wire with an $N$-fold screw symmetry in the basis $\{\uparrow,\downarrow\} \otimes \{A_1, A_2, A_3, \ldots A_N\} \otimes \{p_x, p_y, p_z\}$. The Fourier transform of this Hamiltonian yields the

1

general Bloch Hamiltonian in momentum space

$$\hat{H}_0 = \sum_k \psi_k^\dagger \hat{H}(k) \psi_k, \tag{S1}$$

where $k$ denotes the crystal momentum along the chiral axis ($\mathbf{k} = k\hat{\mathbf{z}}$), and the components of the spinor $\psi_k$ represent electron annihilation operators in the basis of Bloch-summed atomic orbitals. In the first-quantized formalism, the Bloch states are constructed from localized orbitals as $|\chi_{\mu\sigma;k}\rangle = \frac{1}{\sqrt{N}} \sum_\mathbf{R} e^{i\mathbf{k}\cdot(\mathbf{R}+\boldsymbol{\mu})}|\mathbf{R}, \boldsymbol{\mu}, \sigma\rangle$ where $|\mathbf{R}, \boldsymbol{\mu}, \sigma\rangle$ represents an orbital of type $\mu$ with spin $\sigma$, centered at atomic site $\boldsymbol{\mu}$ in the unit cell at $\mathbf{R}$. In the spin-major ordering, the $6N \times 6N$ Bloch Hamiltonian $\hat{H}(k)$ separates into spin blocks

$$\hat{H}(k) = \begin{pmatrix} \hat{H}_{\uparrow\uparrow}(k) & \hat{H}_{\uparrow\downarrow}(k) \\ \hat{H}_{\downarrow\uparrow}(k) & \hat{H}_{\downarrow\downarrow}(k) \end{pmatrix}, \tag{S2}$$

where the first term in Eq. (1) of the main text contributes exclusively to the spin-conserving ($\uparrow\uparrow$, $\downarrow\downarrow$) blocks, while the local SOC term couples both spin-conserving and spin-flip ($\uparrow\downarrow$, $\downarrow\uparrow$) sectors. Since the hopping amplitudes $t_{i\mu,j\nu}$ are spin-independent, their contribution to the spin-conserving block $\hat{H}_{\sigma\sigma}(k)$ can be expressed in terms of the orbital hopping matrix (subspace) between atoms $A_i$ and $A_j$, denoted by $\hat{T}^{ij}$. In the Bloch basis, its components form a $3 \times 3$ matrix (since $\hat{T}^{ij}$ does not act on the spin degree of freedom, we omit the spin label)

$$\begin{aligned} T^{ij}_{\mu\nu}(k) &\doteq \langle \chi_{\mu\sigma;k} | \hat{T}^{ij} | \chi_{\nu\sigma;k} \rangle \\ &= \frac{1}{N} \sum_{\mathbf{R}',\mathbf{R}''} e^{-i\mathbf{k}\cdot(\mathbf{R}''+\boldsymbol{\mu})} e^{i\mathbf{k}\cdot(\mathbf{R}'+\boldsymbol{\nu})} \langle \mathbf{R}'', \boldsymbol{\mu}, \sigma | \hat{T}^{ij} | \mathbf{R}', \boldsymbol{\nu}, \sigma \rangle \\ &= \sum_\mathbf{R} t_{i\mu,j\nu} e^{i\mathbf{k}\cdot(\mathbf{R}-\boldsymbol{\mu}+\boldsymbol{\nu})}, \end{aligned} \tag{S3}$$

where $t_{i\mu,j\nu} = \langle 0, \boldsymbol{\mu}, \sigma | \hat{T}^{ij} | \mathbf{R}, \boldsymbol{\nu}, \sigma \rangle$, explicitly represents the hopping matrix elements, which can be parameterized using standard Slater-Koster integrals.

For an atom $A_i$, the SOC contribution in Eq. (1) in the local orbital-spin sub-basis $\{|p_x \uparrow\rangle, |p_y \uparrow\rangle, |p_z \uparrow\rangle, |p_x \downarrow\rangle, |p_y \downarrow\rangle, |p_z \downarrow\rangle\}$ takes the block form

$$\hat{H}_{\text{SOC}}^{(A_i)} = \lambda_{\text{SOC}} \begin{pmatrix} \hat{\boldsymbol{\sigma}}_{\uparrow\uparrow} \cdot \hat{\boldsymbol{L}}^i & \hat{\boldsymbol{\sigma}}_{\uparrow\downarrow} \cdot \hat{\boldsymbol{L}}^i \\ \hat{\boldsymbol{\sigma}}_{\downarrow\uparrow} \cdot \hat{\boldsymbol{L}}^i & \hat{\boldsymbol{\sigma}}_{\downarrow\downarrow} \cdot \hat{\boldsymbol{L}}^i \end{pmatrix}, \tag{S4}$$

where $\hat{\boldsymbol{L}}^i$ acts in the $p$-orbital subspace of atom $A_i$. Equation S4 can be written explicitly as

$$\hat{H}_{\text{SOC}}^{(A_i)} = \lambda_{\text{SOC}} \begin{pmatrix} \hat{L}_z^i & \hat{L}_x^i - i\hat{L}_y^i \\ \hat{L}_x^i + i\hat{L}_y^i & -\hat{L}_z^i \end{pmatrix}, \tag{S5}$$

with the orbital matrices given by (in $\hbar$ units)

$$\hat{L}_x \doteq \begin{pmatrix} 0 & 0 & 0 \\ 0 & 0 & -i \\ 0 & i & 0 \end{pmatrix}, \quad \hat{L}_y \doteq \begin{pmatrix} 0 & 0 & i \\ 0 & 0 & 0 \\ -i & 0 & 0 \end{pmatrix}, \quad \hat{L}_z \doteq \begin{pmatrix} 0 & -i & 0 \\ i & 0 & 0 \\ 0 & 0 & 0 \end{pmatrix}. \tag{S6}$$

Therefore, in spin-major ordering, the Bloch Hamiltonian blocks between atoms $A_i$ and $A_j$ are given by

$$\begin{aligned} \hat{H}_{\uparrow\uparrow}^{ij}(k) &= \hat{T}^{ij}(k) + \delta_{ij}\, \lambda_{\text{SOC}}\, \hat{L}_z, \\ \hat{H}_{\downarrow\downarrow}^{ij}(k) &= \hat{T}^{ij}(k) - \delta_{ij}\, \lambda_{\text{SOC}}\, \hat{L}_z, \\ \hat{H}_{\uparrow\downarrow}^{ij}(k) &= \delta_{ij}\, \lambda_{\text{SOC}}\, \hat{L}_-, \\ \hat{H}_{\downarrow\uparrow}^{ij}(k) &= \delta_{ij}\, \lambda_{\text{SOC}}\, \hat{L}_+, \end{aligned} \tag{S7}$$

where $\hat{L}_\pm = \hat{L}_x \pm i\hat{L}_y$. Finally, these contributions assemble into the full Bloch Hamiltonian Eq. (S2), where each entry acting on the orbital subspace of the atomic sites. As discussed in the main text, this term breaks the screw symmetry and leads to the opening of energy gaps at specific points along the 1D chiral wire.

In the presence of a rotating electric field $\mathbf{E}(\varphi) = E_0(\cos\varphi\, \hat{\mathbf{x}} + \sin\varphi\, \hat{\mathbf{y}})$, $(\varphi = \omega t)$ the TB

Hamiltonian in Eq. (6) of the main text can be similarly expressed in the reciprocal space as

$$\hat{H}_{\uparrow\uparrow}^{ij}(k,\varphi) = \hat{T}^{ij}(k) + \delta_{ij}\left[\lambda_{\text{SOC}}\,\hat{L}_z - e\,\hat{I}_{3\times 3}^{(\text{orb})}\mathbf{E}(\varphi)\cdot\mathbf{r}_i\right],$$

$$\hat{H}_{\downarrow\downarrow}^{ij}(k,\varphi) = \hat{T}^{ij}(k) - \delta_{ij}\left[\lambda_{\text{SOC}}\,\hat{L}_z + e\,\hat{I}_{3\times 3}^{(\text{orb})}\mathbf{E}(\varphi)\cdot\mathbf{r}_i\right],$$

$$\hat{H}_{\uparrow\downarrow}^{ij}(k,\varphi) = \delta_{ij}\,\lambda_{\text{SOC}}\,\hat{L}_-,$$

$$\hat{H}_{\downarrow\uparrow}^{ij}(k,\varphi) = \delta_{ij}\,\lambda_{\text{SOC}}\,\hat{L}_+.$$

(S8)

Here, $\mathbf{r}_i$ denotes the position of atom $A_i$, and $\hat{I}_{3\times 3}^{(\text{orb})}$ is the identity matrix in the $\{p_x, p_y, p_z\}$ orbital subspace. In the atom-centered TB representation, the position operator acts diagonally in the local orbital basis $|\mathbf{R}_i, \boldsymbol{\mu}, \sigma\rangle$, such that

$$\langle \mathbf{R}_i, \boldsymbol{\mu}, \sigma | \hat{\mathbf{r}} | \mathbf{R}_j, \boldsymbol{\nu}, \sigma' \rangle \simeq \mathbf{r}_i\,\delta_{ij}\delta_{\mu\nu}\delta_{\sigma\sigma'}, \tag{S9}$$

where $\mathbf{r}_i$ denotes the position of the $i$th atomic center. Consequently, from Eq. (S8) the electric-field coupling term in the spin block Hamiltonian takes the compact form

$$-e\,\hat{I}_{2\times 2}^{(\text{spin})} \otimes \hat{I}_{3\times 3}^{(\text{orb})} \otimes \text{diag}\big[\mathbf{E}(\varphi)\cdot\mathbf{r}_1,\, \mathbf{E}(\varphi)\cdot\mathbf{r}_2,\, \ldots,\, \mathbf{E}(\varphi)\cdot\mathbf{r}_N\big] = -e\,\mathbf{E}(\varphi)\cdot\hat{\mathbf{r}}, \tag{S10}$$

where $\hat{\mathbf{r}}$ denotes the atom-centered position operator. Hence, the total spin block Hamiltonian can be compactly written as

$$\hat{H}(k,\varphi) = \hat{I}_{2\times 2}^{(\text{spin})} \otimes \hat{T}(k) + \lambda_{\text{SOC}}\,\hat{\boldsymbol{\sigma}}\cdot\hat{\mathbf{L}} - e\,\mathbf{E}(\varphi)\cdot\hat{\mathbf{r}}. \tag{S11}$$

where $\hat{I}_{2\times 2}^{(\text{spin})}$ the spin-space identity operator and $\hat{\mathbf{r}}$ denotes the total position operator.

# Supplementary Note 2: Fractional translation of Bloch states under $N$-fold screw symmetry

The screw symmetry acts through its translational component $\hat{T}_z(a/N)$, which shifts orbitals along the $z$-axis:

$$\hat{T}_z\left(\tfrac{a}{N}\right)|\mathbf{R},\boldsymbol{\mu},\sigma\rangle = \left|\mathbf{R} + \tfrac{a}{N}\hat{\mathbf{z}},\boldsymbol{\mu},\sigma\right\rangle. \tag{S12}$$

Acting on the Bloch sum gives

$$\hat{T}_z\left(\tfrac{a}{N}\right)|\chi_{\mu\sigma;k}\rangle = \frac{1}{\sqrt{N}}\sum_{\mathbf{R}} e^{i\mathbf{k}\cdot(\mathbf{R}+\boldsymbol{\mu})}\left|\mathbf{R} + \tfrac{a}{N}\hat{\mathbf{z}},\boldsymbol{\mu},\sigma\right\rangle. \tag{S13}$$

Shifting the summation index to $\mathbf{R}' = \mathbf{R} - \tfrac{a}{N}\hat{\mathbf{z}}$ yields

$$\hat{T}_z\left(\tfrac{a}{N}\right)|\chi_{\mu\sigma;k}\rangle = e^{ika/N}|\chi_{\mu\sigma;k}\rangle, \tag{S14}$$

showing that Bloch sums are eigenstates of the fractional translation with eigenvalue $e^{ika/N}$. Consequently, Bloch eigenstates $|\psi_{n\mathbf{k}}\rangle$, being linear combinations of such sums, inherit the same transformation property:

$$\hat{T}_z\left(\tfrac{a}{N}\right)|\psi_{n,k}\rangle = e^{ika/N}|\psi_{n,k}\rangle. \tag{S15}$$

# Supplementary Note 3: Screw symmetry constraints on spin texture

Since in Eq. (1) the Hamiltonian $\hat{H}_0$ respects screw symmetry, it commutes with $\hat{\mathcal{Q}}_N$. Therefore, the Bloch state $|\psi_{n,k}\rangle$ can be chosen as its eigenstate, $\hat{\mathcal{Q}}_N|\psi_{n,k}\rangle = \lambda_N|\psi_{n,k}\rangle$ with

$|\lambda_N|^2 = 1$. The expectation value of a spin component $\hat{S}_i$ ($i = x, y, z$) then satisfies

$$\langle \psi_{n,k} | \hat{S}_i | \psi_{n,k} \rangle = |\lambda_N|^2 \langle \psi_{n,k} | \hat{\mathcal{Q}}_N^\dagger \hat{S}_i \hat{\mathcal{Q}}_N | \psi_{n,k} \rangle = \langle \psi_{n,k} | \hat{\mathcal{Q}}_N^\dagger \hat{S}_i \hat{\mathcal{Q}}_N | \psi_{n,k} \rangle. \tag{S16}$$

For the screw operation $\hat{\mathcal{Q}}_N$, the spatial translation part commutes with the spin operators, so only the spin rotation contributes to the conjugation of $\hat{S}_i$. Therefore, Eq (S16) reduces to $S_i(n,k) = \langle \psi_{n,k} | \hat{R}_z^\dagger \hat{S}_i \hat{R}_z | \psi_{n,k} \rangle$, where a rotation about $z$ by $\theta = 2\pi/N$ is generated by $\hat{R}_z(\theta) = \exp[-\frac{i\theta}{\hbar} \hat{S}_z]$, which acts on spin operators as

$$\begin{aligned} \hat{R}_z^\dagger \hat{S}_x \hat{R}_z &= \hat{S}_x \cos\theta - \hat{S}_y \sin\theta, \\ \hat{R}_z^\dagger \hat{S}_y \hat{R}_z &= \hat{S}_x \sin\theta + \hat{S}_y \cos\theta, \\ \hat{R}_z^\dagger \hat{S}_z \hat{R}_z &= \hat{S}_z. \end{aligned} \tag{S17}$$

The corresponding expectation values therefore satisfy

$$\begin{aligned} S_x(n,k) &= S_x(n,k) \cos\theta + S_y(n,k) \sin\theta, \\ S_y(n,k) &= S_y(n,k) \cos\theta - S_x(n,k) \sin\theta, \\ S_z(n,k) &= S_z(n,k). \end{aligned} \tag{S18}$$

It follows that for $\theta \neq 0$ the only solution at a generic $k$ point is $S_x(n,k) = S_y(n,k) = 0$, whereas $S_z(n,k)$ is left unconstrained by the screw symmetry.

# Supplementary Note 4: Electronic band structure and spin-orbital texture of a trigonal chiral wire

Equation (S7) enables a straightforward numerical implementation and provides a transparent understanding of the TB model, including local SOC effects in the screw-symmetric wire. As an example, we consider the special case of a trigonal chiral wire, illustrated in

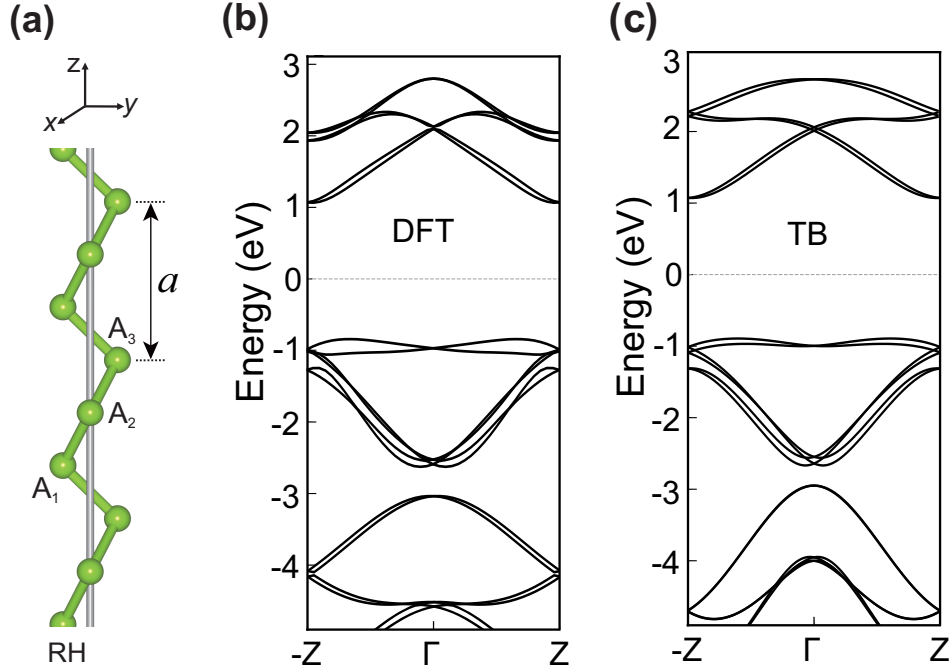

Figure S1: **Tight-binding description of a trigonal chiral wire. a**, Crystal structure of a right-handed trigonal chiral wire with three atoms $A_1$, $A_2$, and $A_3$ per unit cell and lattice constant $a$ along the $z$ axis. **b**, DFT band structure of the isolated right-handed trigonal chiral Se wire with SOC. **c**, Multi-orbital TB band structure of the chiral Se wire including SOC.

Fig. S1a. The unit cell contains three atoms, $A_1$, $A_2$, and $A_3$, and the lattice vector along the $z$ direction is $a$. Using Eqs. (S7) and (S3), the Bloch Hamiltonian blocks in Eq. (S2) are written as

$$\hat{H}_{\uparrow\uparrow}(k) = \hat{T}(k) \ + \ \lambda_{\text{SOC}} \, \text{diag}(\hat{L}_z, \hat{L}_z, \hat{L}_z),$$

$$\hat{H}_{\downarrow\downarrow}(k) = \hat{T}(k) \ - \ \lambda_{\text{SOC}} \, \text{diag}(\hat{L}_z, \hat{L}_z, \hat{L}_z), \quad \text{(S19)}$$

$$\hat{H}_{\uparrow\downarrow}(k) = \lambda_{\text{SOC}} \, \text{diag}(\hat{L}_-, \hat{L}_-, \hat{L}_-) = \hat{H}^\dagger_{\uparrow\downarrow}(k),$$

where the Bloch hopping matrix is

$$\hat{T}(k) = \begin{pmatrix} 0 & h^{12} e^{+ika/3} & h^{13} e^{-ika/3} \\ h^{21} e^{-ika/3} & 0 & h^{23} e^{+ika/3} \\ h^{31} e^{+ika/3} & h^{32} e^{-ika/3} & 0 \end{pmatrix}, \quad \text{(S20)}$$

7

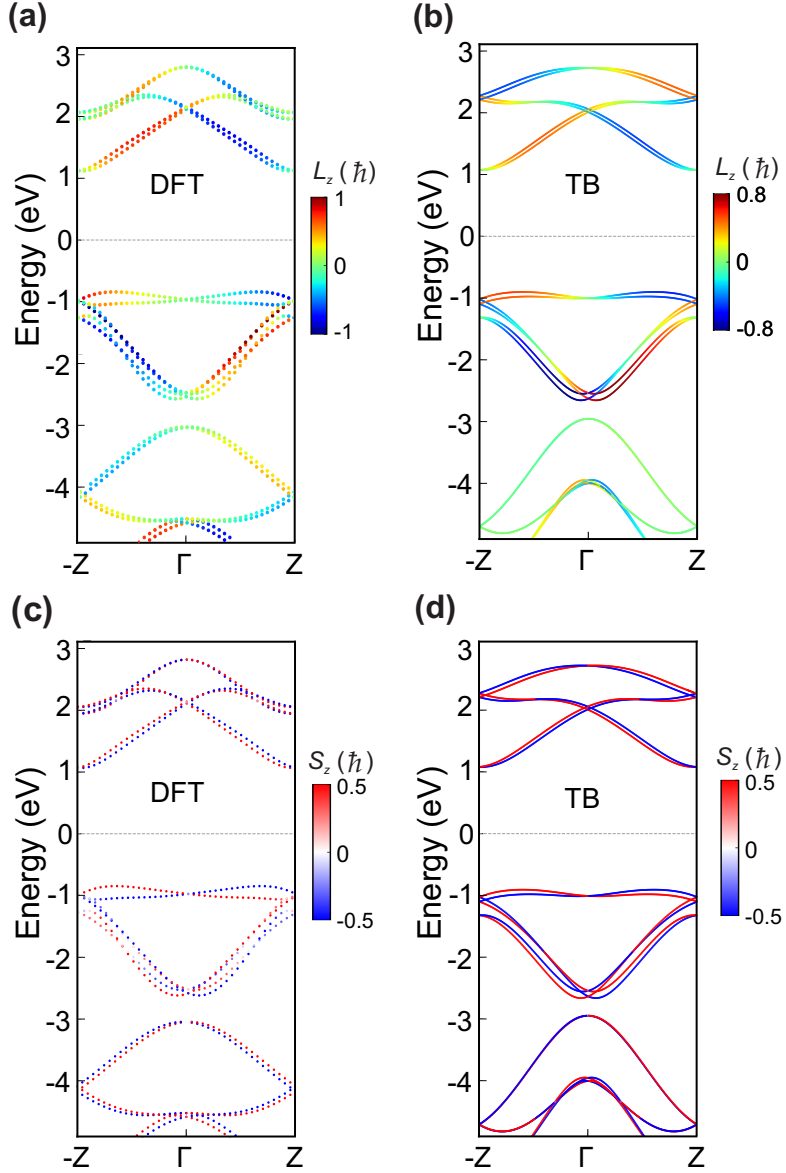

Figure S2: **Orbital and spin angular momentum-resolved textures a,b**, Comparison of the OAM-resolved texture $L_z$ obtained from DFT and TB calculations **c,d**, Comparison of the SAM-resolved texture $S_z$ obtained from DFT and TB calculations.

and the hopping matrix between atoms $A_i$ and $A_j$ is expressed in terms of Slater-Koster parameters for $p$ orbitals as

$$\hat{h}^{ij} = \begin{pmatrix} l_{ij}^2 V_{pp\sigma}^{ij} + (1-l_{ij}^2)V_{pp\pi}^{ij} & l_{ij}m_{ij}(V_{pp\sigma}^{ij} - V_{pp\pi}^{ij}) & l_{ij}n_{ij}(V_{pp\sigma}^{ij} - V_{pp\pi}^{ij}) \\ l_{ij}m_{ij}(V_{pp\sigma}^{ij} - V_{pp\pi}^{ij}) & m_{ij}^2 V_{pp\sigma}^{ij} + (1-m_{ij}^2)V_{pp\pi}^{ij} & m_{ij}n_{ij}(V_{pp\sigma}^{ij} - V_{pp\pi}^{ij}) \\ l_{ij}n_{ij}(V_{pp\sigma}^{ij} - V_{pp\pi}^{ij}) & m_{ij}n_{ij}(V_{pp\sigma}^{ij} - V_{pp\pi}^{ij}) & n_{ij}^2 V_{pp\sigma}^{ij} + (1-n_{ij}^2)V_{pp\pi}^{ij} \end{pmatrix}. \quad (S21)$$

8

Here, $l_{ij}, m_{ij}, n_{ij}$ are the direction cosines of the vector connecting atom $A_i$ to atom $A_j$, and $V_{pp\sigma}^{ij}$ and $V_{pp\pi}^{ij}$ denote the Slater-Koster hopping parameters for the specific pair $(i,j)$.

Using density functional theory, we obtained the SOC-included band structure of an isolated trigonal chiral selenium wire and fitted it to the TB model described above. We consider the hopping terms up to second-nearest neighbors and the fitted parameters are $V_{pp\sigma}^{12} = 2.94$ eV, $V_{pp\pi}^{12} = -0.94194$ eV, $V_{pp\sigma}^{13} = -0.18804$ eV, $V_{pp\pi}^{13} = -0.41555$ eV, and $\lambda_{\text{SOC}} = 0.15$ eV. Figures S1b,c compare the DFT band structure with SOC to the multi-orbital TB results for a right-handed Se wire. The TB model shows very good agreement with the DFT results, capturing the essential features of the band structure with high fidelity.

Figure S2 further demonstrates that the spin-orbital textures obtained from the three-orbital TB model are in good agreement with the corresponding *ab initio* DFT results. For the momentum-resolved spin and orbital angular momentum in the DFT calculations, we use

$$\Lambda_{n,k} = \langle \psi_{n,k} | \hat{\Lambda} | \psi_{n,k} \rangle,$$
$$\hat{S}_z = \frac{\hbar}{2} \begin{pmatrix} 1 & 0 \\ 0 & -1 \end{pmatrix}, \qquad \hat{L}_z = x\hat{p}_y - y\hat{p}_x, \tag{S22}$$

where $\Lambda_{n,k}$ denotes the expectation value of a physical quantity, such as the spin or orbital angular momentum. The operator $\hat{\Lambda}$ represents either the $z$ component of the spin operator ($\hat{S}_z$) or the OAM operator ($\hat{L}_z$).

# Supplementary Note 5: Three-orbital tight-binding model for chiral hydrocarbon $C_{12}H_{12}$

Here, we construct a three-orbital TB Hamiltonian for the spinless chiral hydrocarbon wire $C_{12}H_{12}$ (Fig. S3a). The 1D unit cell contains twelve carbon atoms arranged in a chiral configuration along the wire direction. Each carbon atom contributes three $p$ orbitals which

9

dominate the low-energy electronic structure near the Fermi level. To construct the TB model, we group the twelve carbon atoms into two sets of six sites, labeled $A_1, \ldots, A_6$ and $B_1, \ldots, B_6$, and adopt the orbital basis $\mathcal{B} = \{A_1, \ldots, A_6, B_1, \ldots, B_6\} \otimes \{p_x, p_y, p_z\}$. This basis choice is particularly convenient, as it renders the dominant intersite hopping structure transparent and allows the Hamiltonian to be expressed in the clear block off-diagonal form

$$\hat{H}(k) = \begin{pmatrix} 0 & \hat{H}^{AB}(k) \\ \hat{H}^{BA}(k) & 0 \end{pmatrix} = \sigma_+ \otimes \hat{H}^{AB}(k) + \sigma_- \otimes \hat{H}^{BA}(k),$$

where $\sigma_\pm = (\sigma_x \pm i\sigma_y)/2$ act on the extended sublattices and $\hat{H}^{AB}(k) = \hat{H}^{BA}(k)^\dagger$. The matrix $\hat{H}^{AB}(k)$ is an $18 \times 18$ operator acting in the combined $\{1, \ldots, 6\} \otimes \{p_x, p_y, p_z\}$ subspace. The chiral wire contains two inequivalent types of nearest-neighbor $A$-$B$ bonds as denoted by $C_1$ and $C_2$ (see Fig. S3a). For each of the A-B bonds, we can use the Eq. S21 to construct the corresponding three-orbital subspace hopping matrix. The Slater–Koster parameters associated with the two bond types are obtained by fitting the TB band structure to the *ab initio* results, yielding

$$\text{C1 bonds:} \quad V^{(1)}_{pp\sigma} = 10.74398 \text{ eV}, \quad V^{(1)}_{pp\pi} = -1.98161 \text{ eV},$$
$$\text{C2 bonds:} \quad V^{(2)}_{pp\sigma} = -7.61528 \text{ eV}, \quad V^{(2)}_{pp\pi} = 0.99197 \text{ eV}.$$

The resulting three-orbital TB band structure reproduces the *ab initio* band dispersion near the Fermi energy with excellent accuracy, as shown in Fig. S3b, c.

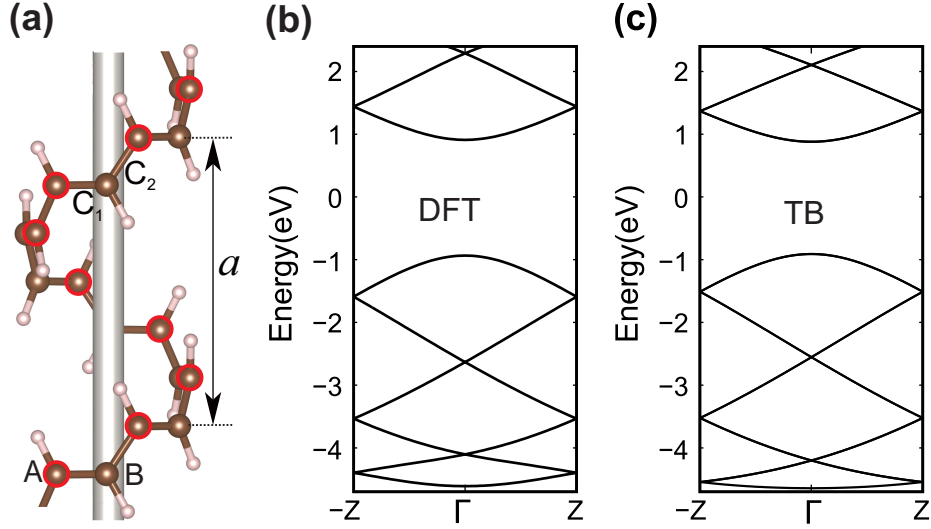

Figure S3: | **Atomic structure and band structure of the helical hydrocarbon wire. a**, Helical hydrocarbon chain $C_{12}H_{12}$ consists of twelve carbon and twelve hydrogen atoms per unit cell. A and B denote the extended sublattice of the carbon chain and $C_1$ and $C_2$ label the two different A-B bonds. **b**, DFT band structure of $C_{12}H_{12}$. **c**, TB band structure of $C_{12}H_{12}$.

# Supplementary Note 6: Field induced orbital tourque during Thouless pumping

From Eq. (26) in the main text and Eq. (S9), the total OAM and position operators take the block-diagonal matrix form (spinless case)

$$\hat{L}_\alpha = \begin{pmatrix} \hat{L}_\alpha^{(A_1)} & 0 & \cdots & 0 \\ 0 & \hat{L}_\alpha^{(A_2)} & \cdots & 0 \\ \vdots & \vdots & \ddots & \vdots \\ 0 & 0 & \cdots & \hat{L}_\alpha^{(A_N)} \end{pmatrix}, \quad \hat{r}_\beta = \begin{pmatrix} \hat{r}_\beta^{(A_1)} & 0 & \cdots & 0 \\ 0 & \hat{r}_\beta^{(A_2)} & \cdots & 0 \\ \vdots & \vdots & \ddots & \vdots \\ 0 & 0 & \cdots & \hat{r}_\beta^{(A_N)} \end{pmatrix}. \quad (S23)$$

For any block-diagonal matrices $X = \mathrm{diag}(X_1, X_2, \ldots, X_N)$ and $Y = \mathrm{diag}(Y_1, Y_2, \ldots, Y_N)$, the commutator acts independently within each block,

$$[X, Y] = \mathrm{diag}([X_1, Y_1], [X_2, Y_2], \ldots, [X_N, Y_N]). \quad (S24)$$

Applying this identity yields

$$[\hat{L}_\alpha, \hat{r}_\beta] = \mathrm{diag}\big([\hat{L}_\alpha^{(A_1)}, \hat{r}_\beta^{(A_1)}], \ldots, [\hat{L}_\alpha^{(A_N)}, \hat{r}_\beta^{(A_N)}]\big). \tag{S25}$$

Using the standard single-atom commutation relation

$$[\hat{L}_\alpha^{(A_i)}, \hat{r}_\beta^{(A_i)}] = i\hbar\, \varepsilon_{\alpha\beta\gamma}\, \hat{r}_\gamma^{(A_i)}, \tag{S26}$$

we obtain

$$[\hat{L}_\alpha, \hat{r}_\beta] = i\hbar\, \varepsilon_{\alpha\beta\gamma} \begin{pmatrix} \hat{r}_\gamma^{(A_1)} & 0 & \cdots & 0 \\ 0 & \hat{r}_\gamma^{(A_2)} & \cdots & 0 \\ \vdots & \vdots & \ddots & \vdots \\ 0 & 0 & \cdots & \hat{r}_\gamma^{(A_N)} \end{pmatrix}$$

$$= i\hbar\, \varepsilon_{\alpha\beta\gamma}\, \hat{r}_\gamma. \tag{S27}$$

Thus, the total OAM operator obeys the same commutation relation as its atomic constituents:

$$[\hat{L}_\alpha,\, \hat{r}_\beta] = i\hbar\, \varepsilon_{\alpha\beta\gamma}\, \hat{r}_\gamma. \tag{S28}$$

Therefore, it directly follows that

$$\frac{i}{\hbar}\Big[-e\,\hat{\mathbf{r}}\cdot\mathbf{E}(\varphi),\, \hat{L}_\alpha\Big] = \frac{ie}{\hbar}\, E_\beta(\varphi)\, [\hat{L}_\alpha, \hat{r}_\beta]$$

$$= -e\, E_\beta(\varphi)\, \varepsilon_{\alpha\beta\gamma}\, \hat{r}_\gamma$$

$$= -e\, \varepsilon_{\alpha\beta\gamma}\, E_\beta(\varphi)\, \hat{r}_\gamma$$

$$= e\, (\hat{\mathbf{r}}\times\mathbf{E}(\varphi))_\alpha, \tag{S29}$$


\end{document}